\documentclass[a4paper,11pt]{article}
\pdfoutput=1 

\usepackage{jheppub} 

\usepackage[T1]{fontenc} 

\usepackage{tikz}
\usetikzlibrary{calc}
\usetikzlibrary{arrows,decorations.pathmorphing,patterns}
\usepackage{graphicx}

\numberwithin{equation}{section}
\usepackage{version}
\usepackage{mathrsfs}

\title{\boldmath Holographic realization of higher-spin Carrollian free fields}

\author{Ethan D'Arcy,}
\author{Arnaud Delfante,}
\author{Stefan Fredenhagen}

\affiliation{Mathematical Physics, Faculty of Physics, University of Vienna,\\
Boltzmanngasse 5, 1090, Vienna, Austria}

\emailAdd{ethan.darcy@univie.ac.at}
\emailAdd{arnaud.delfante@univie.ac.at}
\emailAdd{stefan.fredenhagen@univie.ac.at}

\abstract{We provide a holographic bulk realization of Carrollian free-field structures arising in three-dimensional asymptotically flat (higher-spin) gravity. We construct a class of boundary conditions that generalizes the diagonal gauge of Anti--de Sitter to flat spacetimes. We show that the associated asymptotic symmetries decompose into genuine physical transformations and pure gauge redundancies, the latter being generated by Carrollian screening charges. This structure leads to a bulk-born realization of Carrollian Miura transformations, expressing physical observables in terms of celestial free scalars. Our results establish a concrete link between flat space (higher-spin) gravity and a Carrollian Coulomb gas description, thereby providing a promising route toward the quantization of flat holography.}

\begin{document}

\maketitle
\flushbottom


\section{Introduction} \label{sec. intro}

The AdS/CFT correspondence~\cite{Maldacena:1997re,Witten:1998qj} provides a concrete realization of the holographic principle by relating quantum gravity in Anti--de Sitter (AdS) space to a conformal field theory~(CFT), thereby offering a non-perturbative framework to address conceptual issues in quantum gravity. Since its inception, numerous efforts have aimed at extending this paradigm along various directions, including higher-spin couplings and generalizations to asymptotically flat spacetimes. These are precisely the aspects we investigate in this work. Actually, in both its original formulation and these more recent developments, three-dimensional spacetimes have consistently provided a particularly favorable setting for such investigations. Indeed, Fronsdal fields with spin $s>1$ do not carry local propagating degrees of freedom in three dimensions, which significantly simplifies the analysis. Nevertheless, the corresponding theory remains nontrivial, in that it admits black hole solutions~\cite{Banados:1992wn,Gutperle:2011kf}, and the study of its asymptotic symmetries reveals an enhancement relative to the isometries of the vacuum~\cite{Brown:1986nw,Barnich:2006av,Henneaux:2010xg,Campoleoni:2010zq}. Moreover, in three dimensions, consistent interactions do not require an infinite tower of higher-spin fields~\cite{Aragone:1983sz}. One may instead consider a finite multiplet containing all integer spins between $2$ and some maximal spin~$s$. This feature, together with the absence of radiation, allows one to exploit the topological nature of the interactions and to reformulate the theory as a Chern--Simons~(CS) gauge theory~\cite{Achucarro:1986uwr,Witten:1988hc,Blencowe:1988gj,Bergshoeff:1989ns,Vasiliev:1989re}.

More specifically, in the case of asymptotically AdS$_3$ gravity with Dirichlet boundary conditions, Brown and Henneaux~\cite{Brown:1986nw} showed that the charge algebra consists of a centrally extended double copy of the Virasoro algebra, later identified at the quantum level as the mode algebra of the boundary stress tensor~\cite{Strominger:1997eq}. This result anticipated key features of the AdS/CFT correspondence more than a decade before its formulation. Comparable structures re-emerged nearly twenty-five years later in the context of gravitational coupling to massless higher-spin fields~\cite{Henneaux:2010xg,Campoleoni:2010zq}, formulated in the lowest-weight, or Drinfeld--Sokolov~(DS), gauge~\cite{Drinfeld:1984qv}. In that case, the asymptotic symmetry algebra is given by two copies of a $\mathcal{W}$-algebra, the simplest realization of which, in the spin-three example, is due to Zamolodchikov~\cite{Zamolodchikov:1985wn}, with the Brown--Henneaux central charge. This observation opened the door to higher-spin extensions of the AdS/CFT correspondence and initiated a broad research program. In particular, Gaberdiel and Gopakumar~\cite{Gaberdiel:2010pz} proposed a duality between the three-dimensional higher-spin gravity theory of Prokushkin and Vasiliev~\cite{Prokushkin:1998bq} and an appropriate large-$s$ limit of $\mathcal{W}_s$ minimal model CFTs.

Around the same time, a similar line of developments emerged in the asymptotically flat setting. In this context, Barnich and Comp\`ere showed that the algebra of asymptotic symmetries, obtained under analogous Dirichlet boundary conditions in the Bondi--van der Burg--Metzner--Sachs (BMS) gauge~\cite{Bondi:1960jsa,Sachs:1961zz} and known as the $\mathfrak{bms}_3$ algebra, admits a nontrivial classical central extension of Virasoro type, closely related to its AdS counterpart. Moreover, this algebra turns out to be isomorphic to the two-dimensional Carrollian conformal algebra~\cite{Duval:2014uva}. Taken together, this result, along with the insights gained from the AdS case, provided a conceptual foundation for flat space holography, suggesting that gravity in asymptotically flat spacetimes could admit a dual description in terms of a codimension-one Carrollian conformal field theory (CCFT) living at null infinity (see, e.g.,~\cite{Ruzziconi:2026bix} for a review). This framework was subsequently extended to higher-spin theories~\cite{Afshar:2013vka,Gonzalez:2013oaa} in generalized Bondi gauges. In particular, spin-three gravity in flat space was shown to exhibit an asymptotic symmetry algebra given by a Carrollian analogue of the $\mathcal{W}_3$ algebra. More generally, flat space higher-spin gravity theories, constructed from the corresponding flat higher-spin algebras~\cite{Ammon:2017vwt}, are expected to be governed by Carrollian $\mathcal{W}$-type asymptotic symmetry algebras. These can be understood as resulting from an ultra-relativistic contraction of relativistic $\mathcal{W}$-algebras, extending the Carrollian conformal algebra by higher-spin currents and thereby naturally generalizing Carrollian symmetry structures.

Nevertheless, despite substantial progress in the understanding of asymptotic symmetries and the classical aspects of flat holography, the precise formulation of the dual quantum theory remains largely open. In particular, a controlled framework for quantizing the putative dual Carrollian theory is still lacking, in stark contrast with the AdS/CFT correspondence, where powerful tools from conformal field theory are available. A natural strategy to address this issue is to seek formulations in terms of Carrollian free fields. Indeed, free-field realizations simplify the symplectic structure, lead to tractable Poisson brackets, and provide a promising starting point for quantization. For instance, it has long been known~\cite{Fateev:1987zh} that higher-spin currents in relativistic $\mathcal{W}_s$ algebras can be systematically realized as composite operators built from free bosons, which commute with screening operators via the quantum Miura transformation. This construction is commonly referred to as the Coulomb gas formalism.

Building on~\cite{BALOG199076,Campoleoni:2017xyl}, it has become clear how to implement this Lorentzian boundary construction holographically in the bulk, thereby paving the way for a deeper understanding of the higher-spin/minimal-model duality, in particular from the perspective of a quantization in terms of Darboux coordinates on the phase space of higher-spin fields. More precisely, this approach has been developed in a diagonal gauge of the Chern--Simons theory with higher-spin algebra $\mathfrak{sl}(s,\mathbb{C})$, which describes spins $2,\dots,s$ in Euclidean signature (see, e.g.,~\cite{Campoleoni:2024ced} for a review). Within this gauge, the fundamental variables obey Poisson brackets that reduce to those of free fields. Unlike the Drinfeld--Sokolov gauge, this formulation does not completely fix the gauge symmetry, but instead leaves a finite-dimensional family of residual DS transformations, which should be interpreted as pure gauge symmetries that do not affect the physical state. In particular, it was shown in~\cite{Campoleoni:2017xyl} that the corresponding charges coincide with the classical limits of the screening charges of the Coulomb gas formalism for~$\mathcal{W}_s$ conformal field theories. Moreover, the observables invariant under these residual symmetries are precisely the classical~$\mathcal{W}_s$ currents. In this way, the essential structural ingredients of the Coulomb gas formalism emerge naturally from the bulk higher-spin gravitational theory when formulated in the diagonal gauge.

Recently, it was shown in~\cite{Banerjee:2015kcx,Fredenhagen:2025aqd} how to implement a Carrollian contraction of boundary $\mathcal{W}_s$ currents, expressing them in terms of free-field combinations. These developments opened the way to constructing and analyzing Carrollian representations using free-field techniques. In the present work, we demonstrate that a similar boundary structure naturally arises in asymptotically flat higher-spin gravity, in close analogy with the diagonal gauge of AdS. In particular, the boundary dynamics can be reformulated in terms of celestial free bosons equipped with Carrollian screening charges. This provides an interesting step toward a more systematic understanding of the quantum aspects of (higher-spin) flat holography. It is worth emphasizing that, in order to realize such structures holographically in the bulk, the flat gauge we employ is formally close to that introduced in~\cite{Afshar:2016kjj,Ammon:2017vwt}. However, in the latter, the framework was not formulated so as to fully address the holographic setting in terms of asymptotic symmetries, but was instead primarily focused on (higher-spin) flat space cosmologies with soft hair. Indeed, although a Carrollian Heisenberg structure of the asymptotic symmetries and Miura-like transformations were identified, the boundary conditions were chosen such that all residual symmetries are genuine global symmetries. As a result, the essential holographic link between bulk pure gauge transformations and boundary Carrollian screening charges, together with the physical interpretation in terms of free fields, is not accessible. In this sense, our results fill this gap, even in the spin-two case, and are more closely aligned with the asymptotically AdS constructions of~\cite{BALOG199076,Campoleoni:2017xyl}, where these features play a central role in the emergence of a Coulomb gas description and, ultimately, in the quantization of the theory.

\begin{figure}[ht!]
    \centering
    \includegraphics[width=1\linewidth]{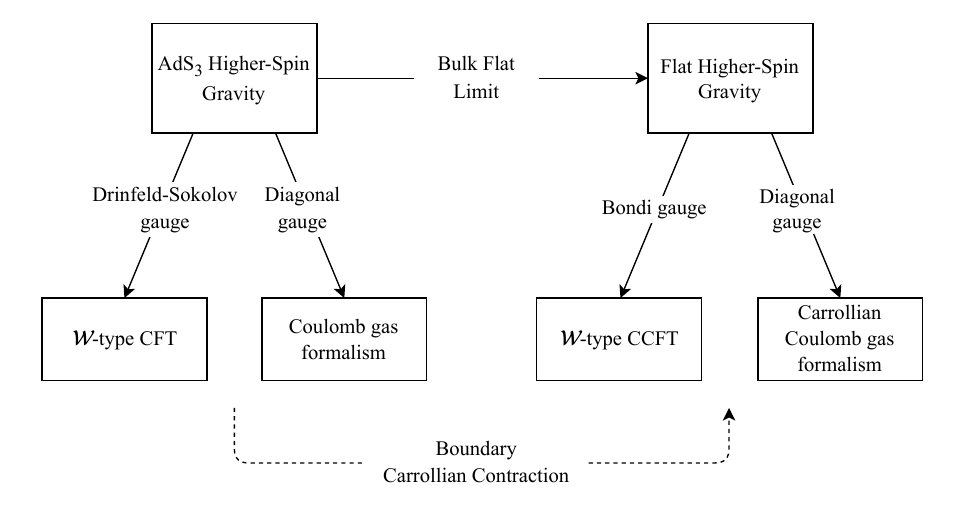}
    \caption{Summary of the AdS and flat gauges, and their holographic interpretations.}
    \label{fig: Introduction Diagram}
\end{figure}

To this end, we begin in Section~\ref{sec. CS} by reviewing the main aspects of the Chern--Simons formulation of three-dimensional higher-spin gravity, and we recast the discussion, for later convenience, within the covariant phase space (CPS) framework of Iyer--Wald (IW)~\cite{Lee:1990nz,Wald:1993nt,Iyer:1994ys}. Then, in Section~\ref{sec. AdS}, we provide in~\ref{subsec. AdS-DS} a review of the spin-two Drinfeld--Sokolov (DS) reduction~\cite{Drinfeld:1984qv,BALOG199076} for asymptotically AdS spacetimes. This corresponds to the standard gauge choice in holography, as it coincides, a posteriori, with the Brown--Henneaux gauge~\cite{Brown:1986nw}. This is followed in~\ref{subsec. AdS-diag} by a review of a second gauge choice, namely the diagonal gauge. In this case, the gauge fixings leave the diagonal directions unconstrained. This analysis was originally initiated in~\cite{BALOG199076,Campoleoni:2017xyl} within the Hamiltonian formalism; here, we revisit it in the CPS framework. This allows us, in particular, to provide a more direct symplectic interpretation of the screening charges as genuine Noether charges associated with pure gauge transformations. This covariant approach proves especially useful for flat and higher-spin extensions, where the corresponding structures are more intricate.

We then turn to the analysis of asymptotically flat spacetimes in Section~\ref{sec. flat}. With this respect, we restart in~\ref{subsec. flat-Bondi} with a review of the Bondi gauge, which can be understood as the flat-space analogue of the DS gauge, before moving in~\ref{subsec. flat-boson} to a regime where genuinely new and particularly interesting features of flat holography emerge. Indeed, following the ``flat from AdS'' philosophy of~\cite{Campoleoni:2023fug}, we generalize the diagonal gauge of AdS to asymptotically flat spacetimes and show that it exhibits a boundary Poisson bracket structure of Carrollian free fields, together with Carrollian screening charges. These ingredients naturally suggest the emergence of a suppositious Carrollian Coulomb gas formalism at the boundary. We then extend this analysis to the coupling to a massless spin-three field in Section~\ref{sec. hs}, within a $\mathfrak{isl}(3)$ Chern--Simons gauge theory, thereby obtaining a holographic bulk realization of the boundary Carrollian Miura transformations studied in~\cite{Fredenhagen:2025aqd}. Although the CS gauge algebras underlying flat (higher-spin) gravity are not semi-simple and therefore do not allow one to single out Cartan directions, we conclude in Section~\ref{sec. beyond} by investigating a representation admitting diagonal generators. This makes it possible to generalize the Carrollian free-field gauge, or Carrollian diagonal gauge, in a systematic algebraic manner to arbitrary values of the spin~$s$ in higher-spin gravitational couplings.


\section{Chern--Simons formalism} \label{sec. CS}

We begin by reviewing the Chern--Simons formulation of three-dimensional (higher-spin) gravity. In this setting, the corresponding couplings are encoded in the action~\cite{Achucarro:1986uwr,Witten:1988hc,Blencowe:1988gj,Bergshoeff:1989ns,Vasiliev:1989re}
\begin{equation} \label{eq. LCS}
    S_\mathrm{CS} = \int_\mathcal{M} \mathrm{L} \, , \qquad \mathrm{L} = \kappa \, \mathrm{Tr} \! \left( \mathcal{A} \wedge \mathrm{d} \mathcal{A} + \frac{2}{3} \mathcal{A} \wedge \mathcal{A} \wedge \mathcal{A} \right) ,
\end{equation}
where $\kappa$ denotes the Chern--Simons level. This level is fixed by Newton’s constant $G$ and the cosmological constant $\Lambda$ of the asymptotic spacetime~$\mathcal{M}$. The field $\mathcal{A}$ is a gauge connection valued in a Lie algebra~$\mathfrak{g}$, whose precise structure is determined by the same data, together with the spectrum of (higher-spin) fields. In the following Sections, we introduce, case by case, a basis of~$\mathfrak{g}$ adapted to the systems under consideration. The corresponding gauge group is denoted by~$G$, and its invariant bilinear form by~$\mathrm{Tr}$.

An arbitrary variation of the Lagrangian~\eqref{eq. LCS} yields
\begin{equation}
    \delta \mathrm{L} = 2\kappa\,\mathrm{Tr}( \mathcal{F} \wedge \delta \mathcal{A} )+ \mathrm{d}\theta \, ,
\end{equation}
where the bulk contribution gives the equations of motion
\begin{equation} \label{eq. eomCS}
    \mathcal{F} = \mathrm{d} \mathcal{A} + \mathcal{A} \wedge \mathcal{A} \approx 0 \, ,
\end{equation}
whereas the boundary contribution is the associated presymplectic potential
\begin{equation} \label{eq. presymp-pot}
    \theta[\mathcal{A}, \delta \mathcal{A}] = - \kappa \, \mathrm{Tr} \! \left( \mathcal{A} \wedge \delta \mathcal{A} \right) .
\end{equation}
In the CPS formalism~\cite{Lee:1990nz,Wald:1993nt,Iyer:1994ys}\footnote{We refer to, e.g.,~\cite{Fiorucci:2021pha,Ciambelli:2022vot,Delfante:2024npo} for pedagogical reviews of these concepts.}, the asymptotic codimension-2 charges are constructed from this last quantity \eqref{eq. presymp-pot}. The corresponding presymplectic two-form is defined as
\begin{equation} \label{eq. presymp-form}
    \omega[\delta \mathcal{A}, \delta \mathcal{A}] = \delta \theta[\mathcal{A}, \delta \mathcal{A}] \, .
\end{equation}
The flatness conditions~\eqref{eq. eomCS} make the absence of local propagating degrees of freedom manifest. Consequently, the dynamics of the theory is entirely encoded in boundary degrees of freedom. Besides, the equations of motion~\eqref{eq. eomCS} are invariant under the finite gauge transformations
\begin{equation} \label{eq. GaugeSym}
    \mathcal{A} \to U^{-1} \mathcal{A} \, U + U^{-1} \mathrm{d} U \, ,
\end{equation}
with $U = \exp(\Lambda) \in G$ and $\Lambda \in \mathfrak{g}$.\footnote{The symbol for the cosmological constant should not be confused with the gauge parameter. The relevant meaning will always be clear from the context.} At the infinitesimal level, this becomes
\begin{equation} \label{eq. InfGaugeSym}
    \delta_\Lambda \mathcal{A} = \mathrm{d}\Lambda + [\mathcal{A},\Lambda] \, .
\end{equation}

The local presymplectic current~\eqref{eq. presymp-form} can be integrated over an arbitrary Cauchy slice~$\Sigma \subset \mathcal{M}$ to define the global presymplectic two-form,
\begin{equation} \label{eq. IWsymp}
    \Omega = \int_\Sigma \omega \, .
\end{equation}
The second Noether theorem then implies that, on shell of the equations of motion~\eqref{eq. eomCS}, the contraction of this form along the gauge transformations~\eqref{eq. InfGaugeSym} gives the so-called surface, or corner, charges of the theory~\cite{Lee:1990nz,Wald:1993nt,Iyer:1994ys}:
\begin{equation} \label{eq. 2ndNoether}
    \omega_\Lambda := \omega[\delta_\Lambda \mathcal{A},\delta \mathcal{A}] = \delta_\Lambda \theta[\mathcal{A},\delta \mathcal{A}] - \delta \theta[\mathcal{A},\delta_\Lambda \mathcal{A}] \approx \mathrm{d}k_\Lambda \, ,
\end{equation}
with associated Iyer--Wald charge density
\begin{equation} \label{eq. kCS}
    k_\Lambda = - 2 \kappa \, \mathrm{Tr} \! \left( \Lambda \, \delta \mathcal{A} \right) .
\end{equation}
Upon integration, one indeed finds the conserved codimension-2 quantities:
\begin{equation} \label{eq. IWsymp-int}
    \Omega_\Lambda = \int_\Sigma \omega_\Lambda \approx \delta H_\Lambda = \int_{\partial \Sigma} k_\Lambda \, .
\end{equation}
This is precisely the covariant form of Hamilton's equation. Actually, the explicit expression~\eqref{eq. kCS} reproduces the surface charge originally obtained in the Hamiltonian formulation of CS theory in~\cite{Banados:1994tn}. Finally, contracting the IW symplectic structure~\eqref{eq. IWsymp} once more yields, on shell, a Poisson bracket on the space of corner charges:
\begin{equation} \label{eq. PB}
    \{H_{\Lambda_1},H_{\Lambda_2}\} \approx \delta_{\Lambda_2} H_{\Lambda_1} \, .
\end{equation}

We now turn to concrete settings, specializing this formal discussion to asymptotically AdS and flat spacetimes, with and without higher-spin couplings.


\section{Asymptotically AdS spacetimes} \label{sec. AdS}

In the presence of a negative cosmological constant, the Einstein--Hilbert action can be recast as a CS theory of the form~\eqref{eq. LCS}, with gauge algebra given by the AdS isometry algebra~$\mathfrak{so}(2,2)$. A convenient basis of generators is
\begin{subequations} \label{eq. so22}
    \begin{align}
        [J_n,J_m] &= (n-m) J_{n+m} \, ,\\
        [J_n,P_m] &= (n-m) P_{n+m} \, ,\\
        [P_n,P_m] &= \frac{1}{\ell^2} (n-m) J_{n+m} \, ,
    \end{align}
\end{subequations}
where $n \in (\pm 1, 0)$, $J_n$ are the Lorentz generators and $P_n$ the translation generators. We have denoted the AdS radius by $\ell$ such that it relates to the cosmological constant via~$\Lambda = -\ell^{-2}$.

Because of the isomorphism $\mathfrak{so}(2,2) \simeq \mathfrak{sl}(2,\mathbb{R}) \oplus \mathfrak{sl}(2,\mathbb{R})$, we can express the AdS$_3$ theory as the subtraction of two $\mathfrak{sl}(2,\mathbb{R})$ Chern--Simons actions~\cite{Achucarro:1986uwr,Witten:1988hc},
\begin{equation} \label{eq. S_split}
    S = S_{\mathrm{CS}}[\mathrm{A}] - S_{\mathrm{CS}}[\bar{\mathrm{A}}] \, ,
\end{equation}
where $\mathrm{A}$ and $\bar{\mathrm{A}}$ are two gauge connections valued in independent copies of $\mathfrak{sl}(2,\mathbb{R})$,
\begin{equation}
    \mathcal{A} = \mathrm{A} + \bar{\mathrm{A}} = \left( A_\mu + \bar{A}_\mu \right) \mathrm{d}x^\mu \, .
\end{equation}
Here $x^\mu$ denotes the bulk spacetime coordinates, and the spacetime metric is mapped via the following definition of the invariant bilinear form:
\begin{equation} \label{eq. gmunu}
    g_{\mu\nu} = {e_\mu}^B \, \eta_{BC} \, {e_\nu}^C = 2 \, \mathrm{Tr} \! \left( e_\mu e_\nu \right) ,
\end{equation}
where $\eta_{BC}$ is the Minkowski metric over the Lorentz indices and $e^B$ is the dreibein. This gravitational interpretation then fixes the CS level to $\kappa = \ell/(16\pi G)$. The two connections are related to the variables of the metric formulation through the combinations
\begin{equation}
    \text{A}^B = \omega^B(e) + \frac{1}{\ell} e^B \, , 
    \qquad 
    \bar{\text{A}}^B = \omega^B(e) - \frac{1}{\ell} e^B \, ,
\end{equation}
where $\omega^B(e)$ denotes the dualized spin connection. We shall use the following basis of~$\mathfrak{sl}(2,\mathbb{R})$:
\begin{equation} \label{eq. sl2}
    [L_n,L_m] = (n-m) L_{n+m} \, , \qquad (n,m = \pm 1,0) \, .
\end{equation}
A similar relation holds for the second gauge copy with corresponding generator $\bar{L}_n$. The corresponding generators are related to those appearing in \eqref{eq. so22} through
\begin{equation} \label{eq. J to L}
    J_n = L_n - \bar{L}_{-n} \, , \qquad P_n = \frac{1}{\ell} \left( L_n + \bar{L}_{-n} \right) .
\end{equation}
The invariance under diffeomorphisms generated by $\xi^\mu$ in the second-order formulation~\eqref{eq. gmunu} is encoded in the first-order formulation through the contraction of the gauge parameters~$\Lambda$ with the inverse dreibein $e_B$,
\begin{equation} \label{eq. ximu}
    \xi^\mu = e^\mu{}_B \, \Lambda^B \, .
\end{equation}

Having introduced the geometric and algebraic aspects, we now turn to the gauge and boundary conditions to be imposed on the fields. One may always profit from the freedom~\eqref{eq. GaugeSym} to cast the CS connections in the radial gauge
\begin{equation} \label{eq. radialgauge}
    \mathrm{A} = b^{-1} \left( \mathrm{a} + \text{d} \right) b \, , \qquad \bar{\mathrm{A}} = \bar{b}^{-1} \left( \bar{\mathrm{a}} + \text{d} \right) \bar{b} \, ,
\end{equation}
such that all radial dependence is carried by the group elements $b$ and $\bar{b}$. This has the convenient consequence that, by cyclicity of the trace, both the action and the associated physical quantities—most notably the symplectic structure and the surface charges—depend only on the radially independent connections $\mathrm{a}$ and $\bar{\mathrm{a}}$,
\begin{equation}
    \mathrm{a} = a_\mu^n \, \mathrm{d}x^\mu L_n \, , \qquad \bar{\mathrm{a}} = \bar{a}_\mu^n \, \mathrm{d}x^\mu \bar{L}_n \, .
\end{equation}
Actually, whenever radial reconstruction will be considered, we shall focus on the standard holographic Fefferman--Graham~\cite{FG1} and Bondi~\cite{Bondi:1960jsa,Sachs:1961zz} gauges. The corresponding group elements are respectively
\begin{equation} \label{eq. bFG}
    \text{FG} : \quad b(z) = \exp\!\left(- \ell \log z \, L_0 \right) , \qquad \bar{b}(z) = b^{-1}(z) \, ,
\end{equation}
and
\begin{equation} \label{eq. bBondi}
    \text{Bondi} : \quad b(r) = \exp\!\left( - \frac{r}{2} P_{-1} \right) .
\end{equation}
In the last two equations \eqref{eq. bFG} and \eqref{eq. bBondi}, the holographic radial coordinate is denoted either by a spacelike coordinate~$z$ or by a null coordinate~$r$, such that the conformal boundary of AdS lies at $z \to 0$ or $r \to \infty$, respectively. Notice that, in the last expression~\eqref{eq. bBondi}, we have assumed to work with the full gauge connection $\mathcal{A}$, since the isomorphic splitting becomes ill-defined in the limit $\ell \to \infty$. As a result, this transformation is the only one among those presented here that admits a smooth flat limit $\Lambda \to 0$. For alternative choices leading to generalized gauges, as well as for analyses based on relaxing the radial gauge condition~\eqref{eq. radialgauge}, we refer, e.g., to~\cite{Grumiller:2016pqb,Grumiller:2017sjh,Campoleoni:2022wmf,Ciambelli:2023ott,Delfante:2025lxn}. We introduce boundary coordinates $(t,\phi)$, where~$t$ denotes time and $\phi \sim \phi + 2\pi$ parametrizes the asymptotic circle. We shall also make frequent use of the light-cone parametrization of the boundary coordinates,
\begin{equation}
    x^\pm = \phi \pm \frac{1}{\ell} t \, .
\end{equation}

Regarding the boundary conditions, one may impose
\begin{equation} \label{eq. varbc}
     \left. A_- \right|_{\partial \mathcal{M}} = 0 \, , \qquad \left. \bar{A}_+ \right|_{\partial \mathcal{M}} = 0 \, ,
\end{equation}
to ensure a well-defined variational principle. Indeed, on shell, the Chern--Simons action reduces to the radial component of the presymplectic potential~\eqref{eq. presymp-pot},
\begin{equation}
    \delta S_\mathrm{CS}[A]
    \approx
    - \frac{\ell}{16 \pi G}
    \int \mathrm{d}x^+ \mathrm{d}x^- \,
    \mathrm{Tr} \! \left(
        A_+ \, \delta A_- - A_- \, \delta A_+
    \right) ,
\end{equation}
and analogously for the second gauge copy. The remaining constraints are chosen such that the fully radial-dependent connections satisfy the asymptotic behavior at conformal infinity
\begin{equation}
    \mathcal{A} - \mathcal{A}_{\mathrm{AdS}} \sim \mathcal{O}(1) \, ,
\end{equation}
where $\mathcal{A}_{\mathrm{AdS}}$ denotes the exact AdS background solution (see, e.g., \cite{Banados:1998gg}). In terms of the radially independent Chern--Simons fields, this requires
\begin{equation} \label{eq. AdS bc}
    \ell^{(1)} = 1 \, , \qquad \bar{\ell}^{(-1)} = - 1 \, ,
\end{equation}
where we have decomposed the fields along the gauge algebra as
\begin{equation}
    a_+(x^+,x^-) = \sum_{n = -1}^{1} \ell^{(n)}(x^+,x^-) L_n \, ,\qquad \bar{a}_-(x^+,x^-) = \sum_{n = -1}^{1} \bar{\ell}^{(n)}(x^+,x^-) \bar{L}_n \, .
\end{equation}

From the Hamiltonian viewpoint, the boundary conditions~\eqref{eq. AdS bc} define first-class constraints, and therefore do not fully fix the gauge. A further gauge fixing may be achieved by imposing additional constraints, for instance $(\ell^{(0)},\bar{\ell}^{(0)}) = (0,0)$ or $(\ell^{(-1)},\bar{\ell}^{(1)}) = (0,0)$. These two choices will be analyzed separately in the following two Subsections \ref{subsec. AdS-DS} and \ref{subsec. AdS-diag}.

\subsection{Drinfeld--Sokolov gauge} \label{subsec. AdS-DS}

We start with the DS gauge. It is defined by the conditions
\begin{equation} \label{eq. AdS DS gauge}
    \ell^{(0)} = 0 \, , \qquad \bar{\ell}^{(0)} = 0 \, ,
\end{equation}
so that the only remaining boundary degrees of freedom can be carried by the components $\ell^{(-1)}$ and $\bar{\ell}^{(1)}$. Solving the equations of motion~\eqref{eq. eomCS}, together with the boundary conditions~\eqref{eq. varbc}, one finds that the solution space is conveniently parametrized by
\begin{equation}
    a_+ = L_1 + \mathscr{L}(x^+) L_{-1} \, , 
    \qquad 
    \bar{a}_- = - \bar{L}_{-1} - \bar{\mathscr{L}}(x^-) \bar{L}_1 \, .
\end{equation}
The gauge freedom is not completely fixed, because of the presence of arbitrary chiral and anti-chiral fields in the physical content, $\mathscr{L}(x^+)$ and $\bar{\mathscr{L}}(x^-)$. Consequently, the solution space still admits residual gauge symmetries, generated by
\begin{subequations} \label{eq. AdS DS lambda}
    \begin{align}
        &\lambda = b^{-1}(z) \left( \varepsilon \, L_1 - \varepsilon' L_0 + \left( \mathscr{L} \varepsilon + \frac{1}{2} \varepsilon'' \right) L_{-1} \right) b(z) \, ,\\
        &\bar{\lambda} = \bar{b}^{-1}(z) \left( \bar{\varepsilon} \, \bar{L}_{-1} - \bar{\varepsilon}' \bar{L}_0 + \left( \bar{\mathscr{L}} \bar{\varepsilon} + \frac{1}{2} \bar{\varepsilon}'' \right) \bar{L}_{1} \right) \bar{b}(z) \, ,
    \end{align}
\end{subequations}
where $\varepsilon = \varepsilon(x^+)$ and $\bar{\varepsilon} = \bar{\varepsilon}(x^-)$ are arbitrary chiral and anti-chiral boundary functions, with the dependence on the radial coordinate $z$ encoded in the group elements $b(z)$ and~$\bar{b}(z)$.\footnote{At this stage, the precise nature of the radial coordinate is not essential, whether it is of Fefferman--Graham or Bondi type. We return to this point below.} Here and in the following, a prime denotes differentiation with respect to the single argument of the corresponding function. The full gauge parameter has been decomposed as~$\Lambda = (\lambda,\bar{\lambda})$. Note that $\varepsilon$ and $\bar{\varepsilon}$ typically encode boundary diffeomorphisms in the asymptotic Killing vectors~\eqref{eq. ximu} of the second-order formulation (see, e.g., \cite{Banados:1998gg}).

Under the modified Lie bracket~\cite{Schwimmer:2008yh,Barnich:2011mi}, which accounts for the field dependence of the generators,
\begin{equation} \label{eq. modifiedbracket}
    \Lambda_{(12)} := [\Lambda_{(1)},\Lambda_{(2)}]_\star = [\Lambda_{(1)},\Lambda_{(2)}] + \delta_{\Lambda_{(1)}} \Lambda_{(2)} - \delta_{\Lambda_{(2)}} \Lambda_{(1)} \, ,
\end{equation}
where $\Lambda_{(1)}$ and $\Lambda_{(2)}$ denote two independent copies of the gauge parameters, the residual symmetries \eqref{eq. AdS DS lambda} are found to close into two copies of the Witt algebra:
\begin{equation} \label{eq. AdS DS resalg}
    \varepsilon_{(12)} = \varepsilon_{(1)} \varepsilon_{(2)}' - \left( 1 \leftrightarrow 2 \right) , 
    \qquad 
    \bar{\varepsilon}_{(12)} = \bar{\varepsilon}_{(1)} \bar{\varepsilon}_{(2)}' - \left( 1 \leftrightarrow 2 \right) .
\end{equation}
The physical fields $\mathscr{L}$ and $\bar{\mathscr{L}}$ transform under these symmetries as the components of an anomalous two-dimensional CFT stress tensor:
\begin{equation} \label{eq. AdS DS transfos}
    \delta_\lambda \mathscr{L} = 2 \, \mathscr{L} \, \varepsilon' + \varepsilon \, \mathscr{L}' + \frac{1}{2} \varepsilon''' \, , 
    \qquad 
    \delta_{\bar{\lambda}} \bar{\mathscr{L}} = - 2 \, \bar{\mathscr{L}} \, \bar{\varepsilon}' - \bar{\varepsilon} \, \bar{\mathscr{L}}' - \frac{1}{2} \bar{\varepsilon}''' \, .
\end{equation}

Applying~\eqref{eq. kCS} and~\eqref{eq. IWsymp-int}, one can compute the surface charges by evaluating the symplectic form along the gauge parameters~\eqref{eq. AdS DS lambda} and integrating over the asymptotic corner, which in the present case is the circle at conformal infinity:
\begin{equation} \label{eq. AdS DS charge}
    H_\Lambda = \frac{\ell}{8 \pi G} \int \mathrm{d}\phi \left( \varepsilon \, \mathscr{L} + \bar{\varepsilon} \, \bar{\mathscr{L}} \right) .
\end{equation}
The $\delta-$integration has been immediate, since we have implicitly assumed $\delta \varepsilon = \delta \bar{\varepsilon} = 0$. More generally, in three dimensions, it has been shown that integrable slicings of gravitational and higher-spin charges can always be constructed~\cite{Ruzziconi:2020wrb,Geiller:2021vpg}, reflecting the absence of local propagating degrees of freedom for fields with spin $s>1$. Besides, all residual gauge parameters enter the asymptotic charges. From the symplectic viewpoint, this indicates that all degenerate directions have been eliminated by gauge fixing; while from the Hamiltonian perspective, \eqref{eq. AdS DS gauge} completes the constraints into a second-class set. Under the Poisson bracket~\eqref{eq. PB}, the charges \eqref{eq. AdS DS charge} furnish a projective representation of the residual symmetry algebra~\eqref{eq. AdS DS resalg} (see, e.g., \cite{Regge:1974zd,Brown:1986ed}),
\begin{equation} \label{eq. chargalg}
    \delta_{\Lambda_{(2)}} H_{\Lambda_{(1)}} = \{H_{\Lambda_{(1)}},H_{\Lambda_{(2)}}\} = H_{[\Lambda_{(1)},\Lambda_{(2)}]_\star} + K_{\Lambda_{(1)}\Lambda_{(2)}} \, ,
\end{equation}
where the 2-cocycle responsible for the central extension reads
\begin{equation}
    K_{\Lambda_{(1)}\Lambda_{(2)}} = \frac{\ell}{8\pi G} \int \mathrm{d}\phi \left( \varepsilon_{(1)} \varepsilon_{(2)}''' + \bar{\varepsilon}_{(1)} \bar{\varepsilon}_{(2)}''' - \left( 1 \leftrightarrow 2 \right) \right) .
\end{equation}
Actually, introducing the Fourier modes
\begin{equation}
    \mathcal{L}_n := H_\Lambda(\varepsilon \sim \mathrm{e}^{\mathrm{i}n x^+}, \bar{\varepsilon} = 0) \, , 
    \qquad 
    \bar{\mathcal{L}}_n := H_\Lambda(\varepsilon = 0, \bar{\varepsilon} \sim \mathrm{e}^{\mathrm{i}n x^-}) \, ,
\end{equation}
one finds that the algebra of charges organizes into two copies of the Virasoro algebra:
\begin{subequations} \label{eq. Virasoro}
    \begin{align}
        \mathrm{i} \{\mathcal{L}_n,\mathcal{L}_m\} &= (n-m) \mathcal{L}_{n+m} + \frac{c}{12} n^3 \delta_{n+m,0} \, ,\\
        \mathrm{i} \{\bar{\mathcal{L}}_n,\bar{\mathcal{L}}_m\} &= (n-m) \bar{\mathcal{L}}_{n+m} + \frac{c}{12} n^3 \delta_{n+m,0} \, ,\\
        \mathrm{i} \{\mathcal{L}_n,\bar{\mathcal{L}}_m\} &= 0 \, ,
    \end{align}
\end{subequations}
with central charge
\begin{equation} \label{eq. BH central charge}
    c = \frac{3\ell}{2G} \, ,
\end{equation}
namely the Brown--Henneaux central charge~\cite{Brown:1986nw}.

To conclude this Subsection, let us collect a few aspects of the radial reconstruction. In the AdS/CFT context, it is natural to adopt the spacelike group element~\eqref{eq. bFG}. At the level of the metric~\eqref{eq. gmunu}, this choice indeed frames the geometry in the standard Fefferman--Graham metric gauge, namely
\begin{equation} \label{eq. FG gauge}
    g_{zz} = \frac{\ell^2}{z^2} \, , \qquad g_{z\pm} = 0 \, .
\end{equation}

We now turn to a different Chern--Simons gauge choice. Its main interest lies in providing a holographic realization of the double copy of the Virasoro algebra in terms of free fields. But we shall also see that it offers a more refined perspective on the holographic reconstruction, by relaxing the metric gauge conditions \eqref{eq. FG gauge}.

\subsection{Diagonal gauge} \label{subsec. AdS-diag}

While still imposing the radial gauge~\eqref{eq. radialgauge} and the boundary conditions~\eqref{eq. varbc} and~\eqref{eq. AdS bc}, the diagonal gauge differs from the Brown--Henneaux choice \eqref{eq. AdS DS gauge} in that it instead constrains the modes
\begin{equation} \label{eq. AdS diag gauge}
    \ell^{(-1)} = 0 \, , \qquad \bar{\ell}^{(1)} = 0 \, ,
\end{equation}
thereby leaving the zero modes $\ell^{(0)}$ and $\bar{\ell}^{(0)}$ arbitrary at the boundary. These modes lie along the Cartan directions of the two copies of the gauge algebra $\mathfrak{sl}(2,\mathbb{R})$. In other words, they are associated with diagonal generators, which is precisely what motivates the name diagonal gauge. On shell, the asymptotic solution space is parametrized by
\begin{equation}
    a_+ = L_1 + \mathscr{J}(x^+) L_0 \, , \qquad \bar{a}_- = - \bar{L}_{-1} + \bar{\mathscr{J}}(x^-) \bar{L}_0 \, ,
\end{equation}
where, for later convenience, we introduce two chiral bosons $\varphi(x^+)$ and $\bar{\varphi}(x^-)$ such that
\begin{equation} \label{eq. AdS J-boson}
    \mathscr{J} = 2 \varphi' \, , \qquad \bar{\mathscr{J}} = 2 \bar{\varphi}' \, .
\end{equation}
The system again admits residual gauge symmetries, here generated by
\begin{subequations}
    \begin{align}
        &\lambda = \mathrm{e}^{2 \varphi} \left( \kappa_1 - \int_{-\infty}^{x^+} \mathrm{d}\bar{x}^+ \, \mathrm{e}^{-2 \varphi} h \right) L_1 + h \, L_0 + \kappa_2 \, \mathrm{e}^{- 2 \varphi} L_{-1} \, ,\\
        &\bar{\lambda} = \mathrm{e}^{- 2 \bar{\varphi}} \left( \bar{\kappa}_1 - \int_{-\infty}^{x^-} \mathrm{d}\bar{x}^- \, \mathrm{e}^{2 \bar{\varphi}} \bar{h} \right) \bar{L}_{-1} + \bar{h} \, \bar{L}_0 + \bar{\kappa}_2 \, \mathrm{e}^{2 \bar{\varphi}} \bar{L}_{1} \, ,
    \end{align}
\end{subequations}
where $h = h(x^+)$ and $\bar{h} = \bar{h}(x^-)$ are arbitrary chiral and anti-chiral boundary functions, and where $(\kappa_{1,2},\bar{\kappa}_{1,2})$ are constants. We now discuss the interpretation of these gauge parameters, whose structure differs markedly from that of the DS case~\eqref{eq. AdS DS lambda}. This becomes even more manifest from the gauge transformations of the physical fields since they now depend not only on a boundary gauge function, but also on a constant:
\begin{equation} \label{eq. AdS diag transfo gauge}
    \delta_\lambda \varphi ' = \frac{1}{2} h ' + \kappa_2 \, \mathrm{e}^{-2 \varphi} \, , 
    \qquad 
    \delta_{\bar{\lambda}} \bar{\varphi} ' = \frac{1}{2} \bar{h} ' + \bar{\kappa}_2 \, \mathrm{e}^{2 \bar{\varphi}} \, .
\end{equation}
In particular we observe that, under the action of the former, the fields transform as free scalars. The corresponding residual symmetry generators form two Abelian copies under the modified bracket~\eqref{eq. modifiedbracket},
\begin{equation}
    h_{(12)} = 0 \, , \qquad \bar{h}_{(12)} = 0 \, ,
\end{equation}
which is a significantly simpler algebraic structure than the double copy of the Witt algebra displayed in~\eqref{eq. AdS DS resalg}.

The precise roles of the gauge parameters are distinguished at the level of the computation of the charges. Repeating the same steps as in the previous Subsection, one finds by direct $\delta$-integration that the asymptotic charges read
\begin{equation} \label{eq. AdS diag charge}
    H_\Lambda = - \frac{\ell}{8 \pi G} \int \mathrm{d}\phi \left( h \, \varphi ' - \bar{h} \, \bar{\varphi} ' \right) ,
\end{equation}
where, once again, we have implicitly assumed $\delta h = \delta \bar{h} = 0$. Actually, at this stage, it is important to pause and discuss the meaning of this result. A first immediate observation is that only the functions $h$ and $\bar{h}$ appear in the asymptotic surface charges. This leads us to conclude that they are the only parameters associated with asymptotic symmetries. Accordingly, the constants $\kappa_{1,2}$ and $\bar{\kappa}_{1,2}$---and in particular $\kappa_2$ and $\bar{\kappa}_2$, since they appear explicitly in the gauge transformations of the physical fields~\eqref{eq. AdS diag transfo gauge}---parametrize degenerate directions of the underlying symplectic structure.\footnote{This also suggests that the constants $\kappa_1$ and $\bar{\kappa}_1$ act as stabilizers on the residual symmetry space.} They are therefore associated with pure gauge symmetries of the system.

A second observation is more subtle, although closely related. It concerns the Iyer--Wald surface charge density~\eqref{eq. kCS}, from which we obtained the charges~\eqref{eq. AdS diag charge}. To better frame the discussion, let us briefly return to some basic notions of the covariant phase space formalism developed in Section \ref{sec. CS}. In this context, the Noether current density is defined as
\begin{equation} \label{eq. jLambda}
    j_\Lambda = \theta[\mathcal{A},\delta_\Lambda \mathcal{A}] - \mathrm{B}_\Lambda \, ,
\end{equation}
where $\theta[\mathcal{A},\delta_\Lambda \mathcal{A}]$ is the presymplectic potential~\eqref{eq. presymp-pot} evaluated along the symmetries and~$\mathrm{B}_\Lambda$ is the boundary term appearing in the gauge variation of the Lagrangian density,
\begin{equation}
    \delta_\Lambda \mathrm{L} = \mathrm{d}\mathrm{B}_\Lambda \, .
\end{equation}
On shell, the Noether current density \eqref{eq. jLambda} coincides with the exterior derivative of the Noether charge density $q_\Lambda$,
\begin{equation}
    j_\Lambda \approx \mathrm{d} q_\Lambda \, ,
\end{equation}
with
\begin{equation} \label{eq. qLambda}
    q_\Lambda = - 2 \kappa \, \mathrm{Tr} \! \left( \Lambda \, \mathcal{A} \right) .
\end{equation}
One may then show that, for an internal gauge symmetry such as in Chern--Simons theory, the Iyer--Wald charge density \eqref{eq. 2ndNoether} is related to the Noether charge density through
\begin{equation} \label{eq. kqqdelta}
    k_\Lambda = \delta q_\Lambda - q_{\delta \Lambda} \, .
\end{equation}
We emphasize once again that it is $k_\Lambda$ that actually determines the asymptotic symmetries. Then, in principle, one wishes to associate with it a Hamiltonian generator $H_\Lambda$ such that
\begin{equation}
    \delta H_\Lambda = \int_{\partial \Sigma} k_\Lambda \, .
\end{equation}
Off shell, one sees that what prevents the Iyer--Wald charge density from being a pure $\delta$-variation is precisely the last term in~\eqref{eq. kqqdelta}, whose origin lies in the possible field dependence of the residual gauge parameters. On shell, however, this is precisely the computation that we succeeded in~\eqref{eq. AdS diag charge}. This ultimately leads us to formulate the second observation, namely a more careful analysis of the meaning of~\eqref{eq. kqqdelta}.

What must be understood from \eqref{eq. kqqdelta} is that the term $q_{\delta \Lambda}$ removes the degenerate directions of the underlying symplectic structure. In this way, $k_\Lambda$ distinguishes, within~$q_\Lambda$, asymptotic symmetries from pure gauge ones. It was the first observation we made. Actually, in standard cases such as the Brown--Henneaux gauge discussed in the previous Subsection \ref{subsec. AdS-DS}, or the Bondi gauge to be studied in the next Section \ref{sec. flat}, the quantity $q_{\delta \Lambda}$ carries no additional physical information beyond the Iyer--Wald charge density, since those setups are based on sets of second-class constraints. In the present case, however, we have seen from the first observation that pure gauge symmetries remain among the residual symmetries. Here, $q_{\delta \Lambda}$ therefore has a physical interpretation distinct from that of $k_\Lambda$. While the functional integration of $k_\Lambda$ yields the Hamiltonian generator of asymptotic symmetries, we thus infer that the functional integration of $q_{\delta \Lambda}$ yields the generator of the residual pure gauge transformations.

Indeed, in the present diagonal gauge, one finds explicitly
\begin{equation} \label{eq. AdS diag qdelta}
    q_{\delta \Lambda} = - \frac{\ell}{4 \pi G} \left( \kappa_2 \, \mathrm{e}^{-2 \varphi} \delta \varphi - \bar{\kappa}_2 \, \mathrm{e}^{2 \bar{\varphi}} \delta \bar{\varphi} \right) \mathrm{d}\phi \, ,
\end{equation}
which is manifestly distinct from~\eqref{eq. AdS diag charge}, since it involves a completely different set of residual gauge parameters. In other words, because the constants $\kappa_2$ and $\bar{\kappa}_2$ appearing in~\eqref{eq. AdS diag qdelta} do not contribute to the asymptotic surface charges~\eqref{eq. AdS diag charge}---that is, $k_\Lambda = 0$ for these parameters---it follows directly from~\eqref{eq. kqqdelta} that
\begin{equation} \label{eq. Noether pure gauge}
    \delta q_\Lambda = q_{\delta \Lambda} \, .
\end{equation}
This means that there exists a functional associated with~\eqref{eq. AdS diag qdelta}, namely a Noether charge that plays the role of generator of the pure gauge symmetries:
\begin{equation} \label{eq. generators pure gauge}
    \delta S_\Lambda = \int_{\partial \Sigma} q_{\delta \Lambda} \, ,
\end{equation}
which in the present case is given explicitly by
\begin{equation} \label{eq. AdS diag screening}
    S_\Lambda = \frac{\ell}{8 \pi G} \int \mathrm{d}\phi \left( \kappa_2 \, \mathrm{e}^{-2 \varphi} + \bar{\kappa}_2 \, \mathrm{e}^{2 \bar{\varphi}} \right) .
\end{equation}
These generators are precisely those identified in~\cite{Campoleoni:2017xyl} by canonical construction from the Poisson brackets that these generators must satisfy on the diagonal-gauge bosons. They were also related there to the screening charges of the classical Coulomb gas formalism. For this reason, we will often refer to them in what follows under that name. Here, however, we have derived them directly from first principles in the CPS formalism. This perspective will be particularly useful later on for higher-spin theories in asymptotically flat spacetimes, where the structures become more involved due to the intrinsic complexity of the asymptotic boundary, as compared to the asymptotically AdS case, and where a boundary theory in terms of a hypothetical Carrollian Coulomb gas remains to be formulated. Note that the above reasoning can even be reversed to show that, for the generators of asymptotic symmetries, $q_{\delta \Lambda} = 0$, and hence
\begin{equation}
    \delta q_\Lambda = k_\Lambda \, ,
\end{equation}
implying that, in this particular three-dimensional setting, the asymptotic charge densities are always functionally integrable. This confirms, and even strengthens, the statements of~\cite{Ruzziconi:2020wrb,Geiller:2021vpg} regarding the integrability of three-dimensional gravity.

What remains to be verified, of course, is that the screening charges~\eqref{eq. AdS diag screening} indeed generate the pure gauge part of the scalar transformations~\eqref{eq. AdS diag transfo gauge}, as suggested by the above Noether charge argument. To establish this explicitly, we need the Poisson bracket structure of the free fields, that is, the symplectic structure reduced by quotienting out the degenerate directions. This follows from the algebra of asymptotic charges. More precisely, one checks that the charges~\eqref{eq. AdS diag charge} furnish, under the bracket~\eqref{eq. PB}, a projective representation of the Abelian residual symmetry algebra~\eqref{eq. chargalg}. This is achieved with the following 2-cocycle:
\begin{equation}
    K_{\Lambda_{(1)}\Lambda_{(2)}} = - \frac{\ell}{16 \pi G} \int \mathrm{d}\phi \left( h_{(1)} h_{(2)}' - \bar{h}_{(1)} \bar{h}_{(2)}' - \left( 1 \leftrightarrow 2 \right) \right) .
\end{equation}
Besides, introducing the Fourier modes
\begin{equation}
    \Phi_n := H_\Lambda(h \sim \mathrm{e}^{\mathrm{i}n x^+}, \bar{h} = 0) \, , 
    \qquad 
    \bar{\Phi}_n := H_\Lambda(h = 0, \bar{h} \sim \mathrm{e}^{\mathrm{i}n x^-}) \, ,
\end{equation}
one finds that the charge algebra organizes into two copies of the Heisenberg algebra:
\begin{equation}
    \mathrm{i} \{\Phi_n,\Phi_m\} = - \frac{c}{12} n \, \delta_{n+m,0} \, , \qquad
    \mathrm{i} \{\bar{\Phi}_n,\bar{\Phi}_m\} = \frac{c}{12} n \, \delta_{n+m,0} \, , \qquad
    \mathrm{i} \{\Phi_n,\bar{\Phi}_m\} = 0 \, ,
\end{equation}
with central charge given by the Brown--Henneaux value~\eqref{eq. BH central charge}. This shows that the reduced boundary symplectic structure is that of two free chiral bosons à la Floreanini--Jackiw~\cite{Floreanini:1987as}:
\begin{equation} \label{eq. AdS Omegabdy}
    \Omega_\mathrm{bdy} = - \frac{\ell}{4 \pi G} \int \mathrm{d}\phi \left( \delta \varphi \wedge \delta \varphi' - \delta \bar{\varphi} \wedge \delta \bar{\varphi}' \right) .
\end{equation}
In particular, it follows that the screening charges correctly act as Noether charges:
\begin{equation}
    \{S_\Lambda,\varphi'\} = \delta_{\kappa_2} \varphi' = \kappa_2 \, \mathrm{e}^{-2 \varphi} \, , 
    \qquad 
    \{S_\Lambda,\bar{\varphi}'\} = \delta_{\bar{\kappa}_2} \bar{\varphi}' = \bar{\kappa}_2 \, \mathrm{e}^{2 \bar{\varphi}} \, .
\end{equation}

A last point to verify along these lines is that these pure gauge generators~\eqref{eq. AdS diag screening} preserve the physical observables $\mathcal{O}$. Since the latter are gauge invariant, one should indeed have
\begin{equation} \label{eq. screening condition Miura}
    \{S_\Lambda,\mathcal{O}\} = 0 \, ,
\end{equation}
thereby confirming that $S_\Lambda$ generate a redundancy rather than a genuine physical symmetry. In the asymptotically AdS case, we reviewed in the previous Subsection \ref{subsec. AdS-DS} that these observables are the components $\mathscr{L}$ and $\bar{\mathscr{L}}$ of an anomalous two-dimensional CFT stress tensor, in the Drinfeld--Sokolov \eqref{eq. AdS DS gauge} gauge. The diagonal gauge \eqref{eq. AdS diag gauge} is related to the latter gauge by
\begin{equation}
    \text{a}_\mathrm{diag} = b^{-1} \, \text{a}_\mathrm{DS} \, b + b^{-1} \, \mathrm{d}b \, , \qquad 
    \bar{\text{a}}_\mathrm{diag} = \bar{b}^{-1} \, \bar{\text{a}}_\mathrm{DS} \, \bar{b} + \bar{b}^{-1} \, \mathrm{d}\bar{b} \, ,
\end{equation}
with
\begin{equation}
    b = \exp \! \left( \varphi' L_{-1} \right) \, , \qquad 
    \bar{b} = \exp \! \left( \bar{\varphi}' \bar{L}_1 \right) \, .
\end{equation}
It follows that we can read the physical observables in the diagonal gauge through the following combinations of the free fields, called Miura transformations \cite{Fateev:1987zh}:
\begin{equation} \label{eq. AdS Miura}
    \mathscr{L} = - \left( \varphi' \right)^2 - \varphi'' \, , \qquad 
    \bar{\mathscr{L}} = - \left( \bar{\varphi}' \right)^2 + \bar{\varphi}'' \, ,
\end{equation}
and one checks explicitly that
\begin{equation}
    \{S_\Lambda,\mathscr{L}\} = 0 \, , \qquad \{S_\Lambda,\bar{\mathscr{L}}\} = 0 \, .
\end{equation}
One can also recover the physical Virasoro transformations~\eqref{eq. AdS DS transfos} by dressing the gauge parameters of the diagonal symplectic slice in a field-dependent manner à la Feigin--Fuchs--Wakimoto~\cite{Feigin:1983rt,Wakimoto:1986gf}, as
\begin{equation}
    \frac{1}{2} h ' + \kappa_2 \, \mathrm{e}^{-2 \varphi} 
    = \varepsilon \, \varphi'' + \varepsilon' \varphi' - \frac{1}{2} \varepsilon'' \, .
\end{equation}
An analogous relation holds for the second copy:
\begin{equation}
    \frac{1}{2} \bar{h} ' + \bar{\kappa}_2 \, \mathrm{e}^{2 \bar{\varphi}} 
    = - \bar{\varepsilon} \, \bar{\varphi}'' - \bar{\varepsilon}' \bar{\varphi}' + \frac{1}{2} \bar{\varepsilon}'' \, .
\end{equation}

We conclude this Subsection with a geometric interpretation, in the second-order formalism, of the diagonal gauge studied above. Reconstructing the spacelike radial coordinate holographically using~\eqref{eq. gmunu} and~\eqref{eq. bFG}, one finds the following metric components:
\begin{equation}
    g_{zz} = \frac{\ell^2}{z^2} \, , \qquad 
    g_{z+} = - \frac{2 \ell^2}{z} \varphi' \, , \qquad 
    g_{z-} = \frac{2 \ell^2}{z} \bar{\varphi}' \, .
\end{equation}
This corresponds to a relaxation of the transverse components with respect to the standard Fefferman--Graham gauge~\eqref{eq. FG gauge}, of the form
\begin{equation} \label{eq. WFG gauge}
    \partial_z \left( \frac{g_{z\pm}}{z} \right) = 0 \, .
\end{equation}
Such a relaxation has been studied in the literature~\cite{Ciambelli:2019bzz,Ciambelli:2023ott} and referred to as the Weyl--Fefferman--Graham gauge, due to its relation with the restoration of radial rescalings through the restoration of a Weyl structure at the boundary.


\section{Asymptotically flat spacetimes} \label{sec. flat}

In this Section, we turn to the case of asymptotically flat spacetimes in three dimensions. When $\Lambda = 0$, gravity can be recast as a Chern--Simons theory~\eqref{eq. LCS} with the Poincar\'e gauge algebra $\mathfrak{iso}(1,2)$. A convenient basis of the corresponding generators can be obtained from the $\mathfrak{so}(2,2)$ AdS algebra~\eqref{eq. so22} in the flat limit $\Lambda \to 0$, or equivalently $\ell \to \infty$:
\begin{subequations}
    \begin{align}
        [J_n,J_m] &= (n-m) J_{n+m} \, ,\\
        [J_n,P_m] &= (n-m) P_{n+m} \, ,\\
        [P_n,P_m] &= 0 \, ,
    \end{align}
\end{subequations}
with $n,m \in \{1,0,-1\}$. Due to the non-semisimple nature of the Minkowski isometry algebra, we prescribe the following invariant bilinear form
\begin{equation}
    \mathrm{Tr}(J_n P_m) = - 2 \begin{pmatrix}
        0 & 0 & 1\\
        0 & - \frac{1}{2} & 0\\
        1 & 0 & 0
    \end{pmatrix} .
\end{equation}
It fixes the CS level to $\kappa = 1/(8\pi G)$ in~\eqref{eq. LCS}. Besides, in contrast with the finite-$\ell$ case, there is no isomorphism allowing for a decomposition into two independent gauge copies. As a result, one must work directly with the full Chern--Simons connection $\mathcal{A}$. In terms of the dreibein and the dualized spin connection, it takes the form
\begin{equation}
    \mathcal{A} = \omega^B J_B + e^B P_B \, .
\end{equation}
In what follows, we again work in the radial gauge~\eqref{eq. radialgauge},
\begin{equation} \label{eq. flat radial gauge}
    \mathcal{A} = b^{-1}(r) \left( \alpha + \mathrm{d} \right) b(r) \, ,
\end{equation}
where we adopt the Bondi radial group element~\eqref{eq. bBondi}, appropriate to null surface, since the Fefferman--Graham choice~\eqref{eq. bFG} is ill-defined in the flat limit. The choice \eqref{eq. flat radial gauge} implies in particular that
\begin{equation}
    \partial_r \, \alpha = 0 \, .
\end{equation}
We also introduce the boundary coordinates $(u,\phi)$ adapted to future null infinity, where the retarded time~$u$ is defined by
\begin{equation}
    u = t - r \, .
\end{equation}

At the level of boundary conditions ensuring a well-defined variational principle, we do not fix any component to vanish, in contrast with~\eqref{eq. varbc}. Rather, since the presymplectic potential involves the combination $\alpha_u \delta \alpha_\phi$, the temporal component $\alpha_u$ is naturally interpreted as a Lagrange multiplier in the Hamiltonian framework, or equivalently as a source from the holographic viewpoint~\cite{Grumiller:2017sjh}. We shall therefore impose additional gauge and boundary conditions only on the angular component $\alpha_\phi$, and their consequences propagate in time solely through the equations of motion. Accordingly, the remaining boundary conditions are chosen such that the fully radial-dependent connections satisfy the asymptotic behavior
\begin{equation}
    \mathcal{A} - \mathcal{A}_{\mathrm{flat}} \sim \mathcal{O}(1) \, ,
\end{equation}
where $\mathcal{A}_{\mathrm{flat}}$ denotes the exact flat background solution. In terms of the radially independent Chern--Simons fields
\begin{equation}
    \alpha = \sum_{n = -1}^{1} j^{(n)}(u,\phi) J_n + \sum_{m = -1}^{1} p^{(m)}(u,\phi) P_m \, ,
\end{equation}
this implies
\begin{equation} \label{eq. flat bc}
    j_\phi^{(1)} = p_u^{(1)} = 1 \, , \qquad p^{(1)}_\phi = j^{(1)}_u = 0 \, .
\end{equation}

These boundary conditions~\eqref{eq. flat bc}, describing an asymptotically flat spacetime, can be obtained from their AdS counterparts~\eqref{eq. AdS bc} in the infinite-$\ell$ limit, using
\begin{equation} \label{eq. flat from AdS - 1}
    \alpha_\phi = \lim_{\ell \to \infty} \left( \alpha_+ + \alpha_- \right) \, , \qquad 
    \alpha_u = \lim_{\ell \to \infty} \frac{1}{\ell} \left( \alpha_+ - \alpha_- \right) \, ,
\end{equation}
together with the change of basis~\eqref{eq. J to L}. As we shall see in the next Subsections \ref{subsec. flat-Bondi} and~\ref{subsec. flat-boson}, residual gauge freedom remains, which we investigate through two gauges obtained by adapting, in a Carrollian fashion, those previously studied in AdS asymptotics.

\subsection{Bondi gauge} \label{subsec. flat-Bondi}

The flat counterpart of the Drinfeld--Sokolov gauge~\eqref{eq. AdS DS gauge}, obtained via the above limit~\eqref{eq. flat from AdS - 1}, is defined by further fixing the modes
\begin{equation} \label{eq. flat Bondi gauge}
    j^{(0)}_\phi = j^{(0)}_u = 0 \, , \qquad p^{(0)}_\phi = p^{(0)}_u = 0 \, .
\end{equation}
This choice coincides retrospectively with the BMS gauge~\cite{Bondi:1960jsa,Sachs:1961zz}, which underlies the construction of a possible BMS/CFT correspondence~\cite{Barnich:2010eb}. On shell, the asymptotic solution space takes the form
\begin{subequations}
    \begin{align}
        \alpha_\phi &= J_1 + M(\phi) \, J_{-1} + \left( L(\phi) + u \, M'(\phi) \right) P_{-1} \, ,\\
        \alpha_u &= P_1 + M(\phi) \, P_{-1} \, ,
    \end{align}
\end{subequations}
where the boundary degrees of freedom are encoded in the Bondi mass aspect $M(\phi)$ and the Bondi angular momentum aspect $L(\phi)$. These quantities arise as the Carrollian limits of the Lorentzian stress-tensor components:
\begin{equation} \label{eq. ML flat limit}
    M = \lim_{\ell \to \infty} \frac{1}{2} \left( \mathscr{L} + \bar{\mathscr{L}} \right) , \qquad 
    L + u \, M' = \lim_{\ell \to \infty} \frac{\ell}{2} \left( \mathscr{L} - \bar{\mathscr{L}} \right) .
\end{equation}

The residual gauge symmetries are generated by the gauge parameter
\begin{equation} \label{eq. flat Lambda}
    \Lambda = b^{-1}(r) \left( \sum_{n=-1}^1 \left( \epsilon^n J_n + \sigma^n P_n \right) \right) b(r) \, ,
\end{equation}
with components
\begin{subequations} \label{eq. flat Bondi resid}
    \begin{align}
        \epsilon^1 &= Y \, ,\\
        \epsilon^0 &= - Y' \, ,\\
        \epsilon^{-1} &= M \, Y + \frac{1}{2} Y'' \, ,\\
        \sigma^1 &= T + u \, Y' \, ,\\
        \sigma^0 &= - T' - u \, Y'' \, ,\\
        \sigma^{-1} &= M \left( T + u \, Y' \right) + Y \left( L + u \, M' \right) + \frac{1}{2} \left( T'' + u \, Y''' \right) .
    \end{align}
\end{subequations}
One can show in the second-order formalism that $T = T(\phi)$ and $Y = Y(\phi)$ parametrize supertranslations and superrotations at future null infinity, respectively, in the asymptotic Killing vectors \eqref{eq. ximu} (see, e.g., \cite{Barnich:2006av}). These parameters are related to the AdS boundary diffeomorphisms via
\begin{equation}
    \varepsilon = Y + \frac{1}{\ell} \left( T + u \, Y' \right) \, , \qquad 
    \bar{\varepsilon} = Y - \frac{1}{\ell} \left( T + u \, Y' \right) \, .
\end{equation}
The double copy of the Witt algebra then contracts to the $\mathfrak{bms}_3$ algebra:
\begin{subequations} \label{eq. bms resid}
    \begin{align}
        T_{(12)} &= Y_{(1)} T_{(2)}' + T_{(1)} Y_{(2)}' - \left( 1 \leftrightarrow 2 \right) \, ,\\
        Y_{(12)} &= Y_{(1)} Y_{(2)}' - \left( 1 \leftrightarrow 2 \right) \, .
    \end{align}
\end{subequations}
Moreover, the Bondi aspects transform under these symmetries as
\begin{subequations} \label{eq. bms transfo}
    \begin{align}
        \delta_\Lambda M &= M' \, Y + 2 \, M \, Y' + \frac{1}{2} Y''' \, ,\\
        \delta_\Lambda L &= L' \, Y + 2 \, L \, Y' + M' \, T + 2 \, M \, T' + \frac{1}{2} T''' \, .
    \end{align}
\end{subequations}

The asymptotic surface charges are integrable, finite, and conserved, and are given by
\begin{equation} \label{eq. flat Bondi charge}
    H_\Lambda = \frac{1}{4 \pi G} \int \mathrm{d}\phi \left( T \, M + Y \, L \right) .
\end{equation}
Since all fields generating the residual symmetries~\eqref{eq. flat Bondi resid} act non-trivially on the phase space through this Hamiltonian, it follows from the symplectic viewpoint that the BMS gauge~\eqref{eq. flat Bondi gauge} defines a set of second-class constraints. Under the Poisson bracket~\eqref{eq. PB}, the corner charges \eqref{eq. flat Bondi charge} furnish a projective representation~\eqref{eq. chargalg} of the residual $\mathfrak{bms}$ algebra~\eqref{eq. bms resid}, with central extension
\begin{equation}
    K_{\Lambda_{(1)}\Lambda_{(2)}} = \frac{1}{8\pi G} \int \mathrm{d}\phi \left( Y_{(1)} T_{(2)}''' + T_{(1)} Y_{(2)}''' - \left( 1 \leftrightarrow 2 \right) \right) .
\end{equation}
Expanding in Fourier modes,
\begin{equation}
    \mathcal{J}_n := H_\Lambda(T = 0,\, Y \sim \mathrm{e}^{\mathrm{i}n\phi}) \, , \qquad 
    \mathcal{P}_n := H_\Lambda(T \sim \mathrm{e}^{\mathrm{i}n\phi},\, Y = 0) \, ,
\end{equation}
one finds the $\mathfrak{bms}_3$ charge algebra
\begin{subequations}
    \begin{align}
        \mathrm{i} \{\mathcal{J}_n,\mathcal{J}_m\} &= (n-m) \mathcal{J}_{n+m} + \frac{c_1}{12} n^3 \delta_{n+m,0} \, ,\\
        \mathrm{i} \{\mathcal{J}_n,\mathcal{P}_m\} &= (n-m) \mathcal{P}_{n+m} + \frac{c_2}{12} n^3 \delta_{n+m,0} \, ,\\
        \mathrm{i} \{\mathcal{P}_n,\mathcal{P}_m\} &= 0 \, ,
    \end{align}
\end{subequations}
with central charges
\begin{equation} \label{eq. flat central charge}
    c_1 = 0 \, , \qquad c_2 = \frac{3}{G} \, .
\end{equation}

To conclude this Subsection, we show that the first-order gauge~\eqref{eq. flat Bondi gauge} indeed reproduces the standard Bondi gauge of the second-order formalism. Reconstructing the metric holographically along the null radial direction using~\eqref{eq. bBondi}, one recovers the Bondi--Sachs conditions:
\begin{equation} \label{eq. Bondi gauge}
    g_{rr} = g_{r\phi} = 0 \, , \qquad g_{ru} = -1 \, , \qquad g_{\phi\phi} = r^2 \, .
\end{equation}

\subsection{Carrollian free field gauge} \label{subsec. flat-boson}

In this Subsection, we present the main spin-two novelties of this work, highlighting interesting aspects for flat holography. The guiding idea is to mimic, in the asymptotically flat setting, the construction achieved in the diagonal gauge for asymptotically AdS spacetimes.

Because the Poincar\'e algebra is not semi-simple, one cannot single out Cartan directions. Nevertheless, it is still possible to introduce two celestial bosons $\psi(\phi)$ and $\bar{\psi}(\phi)$, here defined by
\begin{equation}
    \mathscr{P} = \psi' \, , \qquad \bar{\mathscr{P}} = \bar{\psi}' + u \, \psi'' \, ,
\end{equation}
and such that they arise as Carrollian limits of their Lorentzian counterparts~\eqref{eq. AdS J-boson}:
\begin{equation} \label{eq. P barP flat limit}
    \mathscr{P} = \lim_{\ell \to \infty} \frac{1}{2} \left( \mathscr{J} - \bar{\mathscr{J}} \right) , \qquad 
    \bar{\mathscr{P}} = \lim_{\ell \to \infty} \frac{\ell}{2} \left( \mathscr{J} + \bar{\mathscr{J}} \right) .
\end{equation}
In the spirit of the flat from AdS construction~\cite{Campoleoni:2023fug}, the asymptotic solution space is then given by
\begin{subequations} \label{eq. flat diag bcs}
    \begin{align}
        \alpha_\phi &= J_1 + \psi'(\phi) \, J_0 + \left( \bar{\psi}'(\phi) + u \, \psi''(\phi) \right) P_0 \, ,\\
        \alpha_u &= P_1 + \psi'(\phi) \, P_0 \, .
    \end{align}
\end{subequations}
This corresponds in particular to imposing the constraints
\begin{equation} \label{eq. flat bosonic gauge}
    j^{(-1)}_\phi = j^{(-1)}_u = 0 \, , \qquad p^{(-1)}_\phi = p^{(-1)}_u = 0 \, ,
\end{equation}
so that the boundary degrees of freedom are entirely captured by Carrollian bosons. Notice that the boundary conditions~\eqref{eq. flat diag bcs} differ from the ones considered in~\cite{Afshar:2016kjj} in two respects. First, in that work, the asymptotic fall-offs~\eqref{eq. flat bc} were not imposed. Second, also in that work, chemical potentials were introduced in the temporal component $\alpha_u$. As a result, we are led to say that the boundary conditions~\eqref{eq. flat diag bcs} are better suited for the analysis of asymptotic symmetries in asymptotically flat spacetimes, and thus for their holographic interpretation. For this reason, we adopt them in what follows, while the ones of~\cite{Afshar:2016kjj} seem more appropriate for the study of flat space cosmologies with soft hair.

We now show that the asymptotically flat gauge~\eqref{eq. flat bosonic gauge} shares the same holographic advantage as the AdS diagonal gauge, in the sense that it realizes the Poisson bracket structure in terms of Carrollian free fields. We therefore refer to it as the Carrollian free-field gauge. As we shall see in the final Section~\ref{sec. beyond}, it is nevertheless possible to define a notion of diagonality in terms of the zero modes $J_0$ and $P_0$ of the Poincar\'e algebra, thereby allowing for an alternative interpretation of this gauge as a Carrollian diagonal gauge. In any case, the solution space admits the following residual symmetries:
\begin{subequations} \label{eq. flat bosonic resid}
    \begin{align}
        \epsilon^1 &= \mathrm{e}^{\psi(\phi)} \left( \kappa_1 - \int_{\phi_0}^\phi \mathrm{d}\phi' \, \mathrm{e}^{-\psi(\phi')} \gamma(\phi') \right) ,\\
        \epsilon^0 &= \gamma(\phi) \, ,\\
        \epsilon^{-1} &= \kappa_2 \, \mathrm{e}^{-\psi(\phi)} \, ,\\
        \begin{split}
        \sigma^1 &= \mathrm{e}^{\psi(\phi)} \Bigg\{ \kappa_3 - \int_{\phi_0}^{\phi} \mathrm{d}\phi' \left[ \mathrm{e}^{-\psi(\phi')} \bar{\gamma}(\phi') - \bar{\psi}'(\phi') \left( \kappa_1 - \int_{{\phi_0'}}^{\phi'} \mathrm{d}\phi'' \, \mathrm{e}^{-\psi(\phi'')} \gamma(\phi'') \right) \right]\\
        & \quad + u \, \psi'(\phi) \left[ \kappa_1 - \int_{\phi_0}^{\phi} \mathrm{d}\phi' \, \mathrm{e}^{-\psi(\phi')} \gamma(\phi') \right] \Bigg\} - u \, \gamma(\phi) \, ,
        \end{split}\\
        \sigma^0 &= \bar{\gamma}(\phi) + u \, \gamma'(\phi) \, ,\\
        \sigma^{-1} &= \mathrm{e}^{-\psi(\phi)} \left[ \kappa_4 - \kappa_2 \left( \bar{\psi}(\phi) + u \, \psi'(\phi) \right) \right] .
    \end{align}
\end{subequations}
Here, the modes are decomposed as in~\eqref{eq. flat Lambda}, with $\gamma = \gamma(\phi)$ and $\bar{\gamma} = \bar{\gamma}(\phi)$ arbitrary functions on the celestial circle, and $\kappa_{n=1\ldots4}$ constants. Notice that the residual celestial functions arise as the Carrollian avatars of the Lorentzian scalar gauge parameters, according to
\begin{equation}
    \gamma ' = \lim_{\ell \to \infty} \frac{1}{2} \left( h ' - \bar{h} ' \right) \, , \qquad 
    \bar{\gamma} ' + u \, \gamma '' = \lim_{\ell \to \infty} \frac{\ell}{2} \left( h ' + \bar{h} ' \right) \, .
\end{equation}
The boundary fields then satisfy the transformation laws
\begin{subequations}
    \begin{align}
        \delta_\Lambda \psi'(\phi) &= \gamma'(\phi) + 2 \, \kappa_2 \, \mathrm{e}^{-\psi(\phi)} \, ,\\
        \delta_\Lambda \bar{\psi}'(\phi) &= \bar{\gamma}'(\phi) + 2 \, \mathrm{e}^{-\psi(\phi)} \left( \kappa_4 - \kappa_2 \, \bar{\psi}(\phi) \right) \, .
    \end{align}
\end{subequations}
In particular, they transform as free fields under $\gamma$ and $\bar{\gamma}$, while exhibiting a field-dependent evolution under the constants $\kappa_2$ and $\kappa_4$. Moreover, the residual symmetry algebra generated by~\eqref{eq. flat bosonic resid} takes a remarkably simple form: it is Abelian under the modified bracket~\eqref{eq. modifiedbracket},
\begin{equation}
    \gamma_{(12)} = 0 \, , \qquad \bar{\gamma}_{(12)} = 0 \, .
\end{equation}
We note that only the parameters associated with the zero modes of the Poincar\'e algebra were considered in~\cite{Afshar:2016kjj}, leaving aside the discussion of pure gauge symmetries.

The computation of the asymptotic corner charges again provides a clear physical splitting between the parameters generating the gauge symmetries. Actually, one finds the Hamiltonian
\begin{equation} \label{eq. flat bosonic charge}
    H_\Lambda = - \frac{1}{8 \pi G} \int \mathrm{d}\phi \left( \gamma(\phi) \, \bar{\psi}'(\phi) + \bar{\gamma}(\phi) \, \psi'(\phi) \right) .
\end{equation}
Thus, this makes it clear that the celestial functions $\gamma$ and $\bar{\gamma}$ generate the asymptotic symmetries, while the constants $\kappa_2$ and $\kappa_4$ parametrize degenerate directions of the underlying symplectic structure.\footnote{We observe that the remaining constants stabilize the residual symmetries, as in the AdS case.} Indeed, following the same steps of the covariant phase space formalism as in the discussion below~\eqref{eq. AdS diag charge} in Subsection~\ref{subsec. AdS-diag}, one obtains
\begin{equation} \label{eq. flat diag qdeltaLambda}
    q_{\delta \Lambda} = - \frac{\mathrm{e}^{-\psi}}{4 \pi G} \left( \kappa_2 \, \delta \bar{\psi} + \left( \kappa_4 - \kappa_2 \,\bar{\psi} \right) \delta \psi \right) \mathrm{d}\phi \, .
\end{equation}
We can then apply~\eqref{eq. generators pure gauge} to extract, by functional integration, the generators of pure gauge symmetries, which we refer to, by analogy, as Carrollian screening charges. This procedure is justified by the fact that the key assumption underlying~\eqref{eq. generators pure gauge} is satisfied, namely the existence of a clear separation between $k_\Lambda$ and $q_{\delta \Lambda}$ in terms of the parameters entering the Noether charge~$q_\Lambda$. Performing the $\delta-$integration of \eqref{eq. flat diag qdeltaLambda}, one finds
\begin{equation} \label{eq. flat screening}
    S_{\kappa_2} = - \frac{\kappa_2}{4\pi G} \int \mathrm{d}\phi \, \mathrm{e}^{-\psi(\phi)} \bar{\psi}(\phi) \, , \qquad 
    S_{\kappa_4} = \frac{\kappa_4}{4\pi G} \int \mathrm{d}\phi \, \mathrm{e}^{-\psi(\phi)} \, .
\end{equation}

One may again verify that these screening charges \eqref{eq. flat screening} generate the correct Poisson brackets. To this end, we need the algebraic structure satisfied by the celestial bosons. Under~\eqref{eq. PB}, the asymptotic charges~\eqref{eq. flat bosonic charge} close as in~\eqref{eq. chargalg}, with
\begin{equation}
    K_{\Lambda_{(1)}\Lambda_{(2)}} = - \frac{1}{8\pi G} \int \mathrm{d}\phi \left( \gamma_{(1)} \bar{\gamma}_{(2)}' - \left( 1 \leftrightarrow 2 \right) \right) .
\end{equation}
Expanding as
\begin{equation}
    \Psi_n := H_\Lambda(\gamma = 0,\, \bar{\gamma} \sim \mathrm{e}^{\mathrm{i}n\phi}) \, , \qquad 
    \bar{\Psi}_n := H_\Lambda(\gamma \sim \mathrm{e}^{\mathrm{i}n\phi},\, \bar{\gamma} = 0) \, ,
\end{equation}
one finds that the asymptotic charge algebra takes the form
\begin{equation} \label{eq. flat bosonic charge algebra}
    \mathrm{i} \{\Psi_n,\Psi_m\} = 0 \, , \qquad 
    \mathrm{i} \{\Psi_n,\bar{\Psi}_m\} = - \frac{c_2}{6} n \, \delta_{n+m,0} \, , \qquad 
    \mathrm{i} \{\bar{\Psi}_n,\bar{\Psi}_m\} = 0 \, ,
\end{equation}
with central charge given by the BMS value~\eqref{eq. flat central charge}. We thus infer that the reduced boundary symplectic structure is the ultra-relativistic counterpart of~\eqref{eq. AdS Omegabdy}, that is, the one of a Carrollian scalar:
\begin{equation}
    \Omega_\mathrm{bdy} = - \frac{1}{8 \pi G} \int \mathrm{d}\phi \, \delta \bar{\psi} ' \wedge \delta \psi \, .
\end{equation}
This leads us to define that the dual theory associated with the gauge described in this subsection is governed by a Carrollian Coulomb gas. By analogy with the Lorentzian case, we prescribe that such a system involves fields transforming as free fields, admits a scalar symplectic structure, and possesses screening charges generating pure gauge transformations. Indeed, with this structure in hand, one directly verifies their interpretation as Noether charges:
\begin{subequations}
    \begin{align}
        \{S_{\kappa_2,\kappa_4},\psi'\} &= \delta_{\kappa_2,\kappa_4} \psi' = 2 \, \kappa_2 \, \mathrm{e}^{-\psi(\phi)} \, ,\\
        \{S_{\kappa_2,\kappa_4},\bar{\psi}'\} &= \delta_{\kappa_2,\kappa_4} \bar{\psi}' = 2 \, \mathrm{e}^{-\psi(\phi)} \left( \kappa_4 - \kappa_2 \, \bar{\psi}(\phi) \right) .
    \end{align}
\end{subequations}
It could be interesting to determine whether such a scalar theory corresponds off shell to an electric or magnetic Carrollian limit of the Floreanini--Jackiw chiral bosonic action. Although these two limits are expected to be classically equivalent on shell, they may differ at the quantum level. We leave this question, as well as the deeper physical interpretation of a Carrollian Coulomb gas, for future investigation.

Besides, the Carrollian screening charges allow us to determine, under the condition~\eqref{eq. screening condition Miura}, which combinations of the Carrollian free fields correspond to genuine physical observables, by quotienting out the gauge orbits associated with the degenerate directions of the symplectic structure. One finds that they are given by
\begin{subequations} \label{eq. flat Carroll Miura}
    \begin{align}
        M(\phi) &= - \frac{1}{4} \left( \psi'(\phi) \right)^2 - \frac{1}{2} \psi''(\phi) \, ,\\
        L(\phi) &= - \frac{1}{2} \psi'(\phi) \bar{\psi}'(\phi) - \frac{1}{2} \bar{\psi}''(\phi) \, ,
    \end{align}
\end{subequations}
such that, indeed,
\begin{equation}
    \delta_{\kappa_2,\kappa_4} M = 0 \, , \qquad \delta_{\kappa_2,\kappa_4} L = 0 \, .
\end{equation}
These are referred to as Carrollian Miura transformations and were originally identified from a purely boundary $\mathfrak{bms}_3$ field-theoretic perspective in~\cite{Banerjee:2015kcx}. Here, we have derived them thoroughly holographically from the bulk of three-dimensional asymptotically flat spacetimes and provided them with a physical interpretation. We note, however, that they were mentioned in the distinct cosmological context of \cite{Ammon:2017vwt}. Of course, one can verify that~\eqref{eq. flat Carroll Miura} matches the smooth flat limit of the Lorentzian Miura transformations under the prescriptions~\eqref{eq. ML flat limit} and~\eqref{eq. P barP flat limit}. Actually, we have chosen to label the physical observables in the same way as the Bondi mass and angular momentum aspects in the BMS gauge~\eqref{eq. flat Bondi gauge}, since they are related through the gauge transformation
\begin{equation}
    \alpha_\mathrm{diag} = b^{-1} \left( \alpha_\mathrm{BMS} + \mathrm{d} \right) b \, ,
\end{equation}
with
\begin{equation}
    b = \exp \! \left( \frac{1}{2} \psi' J_{-1} + \frac{1}{2} \left( \bar{\psi}' + u \, \psi'' \right) P_{-1} \right) .
\end{equation}
The Bondi~\eqref{eq. flat Bondi gauge} and Carrollian free-field~\eqref{eq. flat bosonic gauge} gauges thus coincide exactly when the conditions~\eqref{eq. flat Carroll Miura} are satisfied. One may also reconstruct the physical $\mathfrak{bms}_3$ transformations~\eqref{eq. bms transfo} starting from the diagonal symplectic slice, by performing a field-dependent redefinition of the gauge parameters à la Feigin--Fuchs--Wakimoto~\cite{Feigin:1983rt,Wakimoto:1986gf}. This yields
\begin{subequations}
    \begin{align}
        &\gamma' + 2 \, \kappa_2 \, \text{e}^{-\psi} = Y \psi'' + Y' \psi' - Y'' \, ,\\
        &\bar{\gamma}' + 2 \left( \kappa_4 - \kappa_2 \bar{\psi} \right) \text{e}^{-\psi} = Y \bar{\psi}'' + Y' \bar{\psi}' + T \psi'' + T' \psi' - T'' \, .
    \end{align}
\end{subequations}

To conclude this Subsection, let us again provide several geometric interpretations of the gauge analyzed above. Reconstructing the metric holographically via the null radial group element~\eqref{eq. bBondi}, one obtains
\begin{equation}
    g_{rr} = g_{r\phi} = 0 \, , \qquad g_{ru} = -1 \, , \qquad g_{\phi\phi} = \left( r + \bar{\psi}' + u \, \psi'' \right)^2 \, .
\end{equation}
This reveals a relaxation of the purely angular component with respect to the standard Bondi gauge~\eqref{eq. Bondi gauge}, of the form
\begin{equation} \label{eq. Bondi Weyl gauge}
    \partial_r \left( \frac{g_{\phi\phi}}{r^2} \right) = \partial_r W \, , \qquad 
    W = \frac{2}{r} \left( \bar{\psi}' + u \, \psi'' \right) + \frac{1}{r^2} \left( \bar{\psi}' + u \, \psi'' \right)^2 \, .
\end{equation}
This is precisely of the type of generalized Bondi--Sachs condition studied in~\cite{Geiller:2021vpg}. It is referred to as the Bondi--Weyl gauge due to its relation to Weyl rescalings, in close analogy with the Weyl--Fefferman--Graham gauge~\eqref{eq. WFG gauge} of AdS.


\section{Spin three free-field Carroll holography} \label{sec. hs}

In this Section, we investigate the coupling of higher-spin fields to asymptotically flat spacetimes and show that a structure analogous to the spin-two case can be realized holographically, that is, an asymptotic symmetry algebra formulated in terms of Carrollian free fields.

For later convenience, we begin with a brief review of the higher-spin free-field construction from the bulk perspective in the presence of a negative cosmological constant. We assume a spectrum based on the gauge algebra $\mathfrak{sl}(s,\mathbb{R}) \oplus \mathfrak{sl}(s,\mathbb{R})$, in which the AdS isometry algebra $\mathfrak{so}(2,2)$ is embedded diagonally. Alternative choices of embedding may lead to different spectra~\cite{Campoleoni:2024ced}. For concreteness, and to avoid overly lengthy expressions, we focus on the case $s=3$ in this Section. The formal generalization to arbitrary $s$ will be implemented algebraically in Section~\ref{sec. beyond}. In the case of $\mathfrak{sl}(3,\mathbb{R})$, a convenient basis of generators is given by
\begin{subequations} \label{eq. sl3basis}
    \begin{align}
    &[L_i,L_j] = \left(i-j\right) L_{i+j} \, ,\\
    &[L_i,W_n] = \left(2i-n\right) W_{i+n} \, ,\\
    &[W_n,W_m] = - \frac{1}{3} \left(n-m\right) \left(2n^2 + 2m^2 - n m - 8\right) L_{n+m} \, ,
    \end{align}
\end{subequations}
where $-1 \leq i,j \leq 1$ and $-2 \leq n,m \leq 2$, and similarly for the second copy. The analysis of asymptotic symmetries in such higher-spin Chern--Simons theories was initiated in~\cite{Henneaux:2010xg,Campoleoni:2010zq}, in the spin-three Drinfeld--Sokolov gauge (see~\eqref{eq. AdS DS gauge} for comparison with the spin-two case),
\begin{subequations} \label{eq. W3 asymptotic sol space}
    \begin{align}
        a_+ &= L_1 + \mathscr{L}(x^+) L_{-1} + \mathscr{W}(x^+) W_{-2} \, ,  &a_- = 0 \, ,\\
        \bar{a}_- &= - \bar{L}_{-1} - \bar{\mathscr{L}}(x^-) \bar{L}_{1} - \bar{\mathscr{W}}(x^-) \bar{W}_{2} \, , &\bar{a}_+ = 0 \, ,
    \end{align}
\end{subequations}
where it was shown that the asymptotic symmetry algebra consists of two copies of the Zamolodchikov $\mathcal{W}_3$ algebra~\cite{Zamolodchikov:1985wn}, with Brown--Henneaux central charge~\eqref{eq. BH central charge}.

This generalization was subsequently extended to asymptotically flat spacetimes~\cite{Afshar:2013vka,Gonzalez:2013oaa} by identifying the relevant features through a smooth flat limit of the asymptotic solution space~\eqref{eq. W3 asymptotic sol space}. At the algebraic level, this procedure is again considerably simpler than at the geometric level and leads naturally to a Chern--Simons theory with a $\mathfrak{isl}(3)$ gauge algebra. The corresponding generators can then be obtained as descendants of their $\mathfrak{sl}(3,\mathbb{R}) \oplus \mathfrak{sl}(3,\mathbb{R})$ counterparts \eqref{eq. sl3basis} through the limits
\begin{subequations}
    \begin{align}
        &J_i = \lim_{\ell \to \infty} \left( L_i - \bar{L}_{-i} \right) , &&P_i = \lim_{\ell \to \infty} \frac{1}{\ell} \left( L_i + \bar{L}_{-i} \right) ,\\
        &U_n = \lim_{\ell \to \infty} \left( W_n - \bar{W}_{-n} \right) , &&V_n = \lim_{\ell \to \infty} \frac{1}{\ell} \left( W_n + \bar{W}_{-n} \right) ,
    \end{align}
\end{subequations}
which yield the following convenient basis:
\begin{subequations} \label{eq. isl3basis}
    \begin{align}
        [J_i,J_j] &= \left(i-j\right) J_{i+j} \, ,\\
        [J_i,P_j] &= \left(i-j\right) P_{i+j} \, ,\\
        [J_i,U_n] &= \left(2i-n\right) U_{i+n} \, ,\\
        [J_i,V_n] &= \left(2i-n\right) V_{i+n} \, ,\\
        [P_i,U_n] &= \left(2i-n\right) V_{i+n} \, ,\\
        [U_n,U_m] &= -\frac{1}{3} \left(n-m\right) \left(2n^2+2m^2-nm-8\right) J_{n+m} \, ,\\
        [U_n,V_m] &= -\frac{1}{3} \left(n-m\right) \left(2n^2+2m^2-nm-8\right) P_{n+m} \, .
    \end{align}
\end{subequations}
We also introduce the following invariant bilinear form
\begin{equation} \label{eq:isl3bilinform}
    \mathrm{Tr}(J_i P_j) = -2 \begin{pmatrix}
    0 & 0 & 1\\
    0 & -\frac{1}{2} & 0\\
    1 & 0 & 0
    \end{pmatrix} , \qquad
    \mathrm{Tr}(U_nV_m) = \begin{pmatrix}
    0 & 0 & 0 & 0 & 8\\
    0 & 0 & 0 & -2 & 0\\
    0 & 0 & \frac{4}{3} & 0 & 0\\
    0 & -2 & 0 & 0 & 0\\
    8 & 0 & 0 & 0 & 0
    \end{pmatrix} .
\end{equation}

In turn, the AdS fields exhibit the following large-$\ell$ behavior:
\begin{subequations}
    \begin{align}
        &M_{(L)} = \lim_{\ell \to \infty} \frac{1}{2} \left( \mathscr{L} + \bar{\mathscr{L}} \right) , &&L_{(L)} + u \, M_{(L)}' = \lim_{\ell \to \infty} \frac{\ell}{2} \left( \mathscr{L} - \bar{\mathscr{L}} \right) ,\\
        &M_{(W)} = \lim_{\ell \to \infty} \frac{1}{2} \left( \mathscr{W} + \bar{\mathscr{W}} \right) , &&L_{(W)} + u \, M_{(W)}' = \lim_{\ell \to \infty} \frac{\ell}{2} \left( \mathscr{W} - \bar{\mathscr{W}} \right) ,
    \end{align}
\end{subequations}
and the solution space for vanishing cosmological constant is then given by the spin-three Bondi gauge (see \eqref{eq. flat Bondi gauge} for comparison with the pure gravity case)
\begin{subequations} \label{eq. hs Bondi gauge}
    \begin{align}
        \begin{split}
        \alpha_\phi &= J_1 + M_{(L)}(\phi) \, J_{-1} + \left( L_{(L)}(\phi) + u \, M_{(L)}'(\phi) \right) P_{-1}\\
        &\quad + M_{(W)}(\phi) \, U_{-2} + \left( L_{(W)}(\phi) + u \, M_{(W)}'(\phi) \right) V_{-2} \, ,
        \end{split}\\
        \alpha_u &= P_1 + M_{(L)}(\phi) \, P_{-1} + M_{(W)}(\phi) \, V_{-2} \, ,
    \end{align}
\end{subequations}
where the boundary degrees of freedom are encoded in the Bondi mass and angular momentum aspects, $M_{(L)}$ and $L_{(L)}$, together with their higher-spin generalizations, $M_{(W)}$ and $L_{(W)}$~\cite{Campoleoni:2017mbt,Campoleoni:2017qot,Campoleoni:2020ejn,Campoleoni:2025bhn}. It was then shown in~\cite{Afshar:2013vka,Gonzalez:2013oaa} that this construction yields the spin-three extension of the $\mathfrak{bms}_3$ charge algebra, which can be understood as the Carrollian contraction of two copies of the $\mathcal{W}_3$ algebra~\cite{Campoleoni:2016vsh}.

In what follows, we shall not pursue a detailed analysis of the standard higher-spin gauges introduced above. Rather, we will focus on the new results and structures underlying the holographic realization of higher-spin Carrollian free fields. To this end, we will adopt the same conventions, gauge choices, and boundary conditions as those used for asymptotically flat spacetimes at the beginning of Section~\ref{sec. flat}, but working here within the higher-spin algebra~\eqref{eq. isl3basis}. We will then follow the same flat from AdS strategy~\cite{Campoleoni:2023fug} as before: while the Bondi gauge arises from the Drinfeld--Sokolov gauge, we shall construct the Carrollian free field gauge through Laurent expansion of the diagonal gauge in AdS \cite{Campoleoni:2017xyl}.

\subsection{Flat from AdS} \label{subsec. hs-flat}

For $\Lambda < 0$, the Lie algebra $\mathfrak{sl}(3,\mathbb{R}) \oplus \mathfrak{sl}(3,\mathbb{R})$ retains Cartan directions along the zero modes of~\eqref{eq. sl3basis}. A higher-spin generalization of the diagonal gauge was then proposed in~\cite{Campoleoni:2017xyl}, in which the boundary degrees of freedom are organized along these generators:
\begin{subequations}
    \begin{align}
        &a_+ = L_1 + \mathscr{J}_{(L)}(x^+) L_0 + \mathscr{J}_{(W)}(x^+) W_0 \, , &&a_- = 0 \, ,\\
        &\bar{a}_- = - \bar{L}_{-1} + \bar{\mathscr{J}}_{(L)}(x^-) \bar{L}_0 + \bar{\mathscr{J}}_{(W)}(x^-) \bar{W}_0 \, , &&\bar{a}_+ = 0 \, ,
    \end{align}
\end{subequations}
Besides, upon introducing four chiral bosons via
\begin{equation} \label{eq. s=3 bosons}
    \mathscr{J}_{(L,W)} = 2 \varphi_{(L,W)}' \, , \qquad 
    \bar{\mathscr{J}}_{(L,W)} = 2 \bar{\varphi}_{(L,W)}' \, ,
\end{equation}
it was shown in~\cite{Campoleoni:2017xyl} that the double Zamolodchikov asymptotic symmetry algebra is now realized in terms of Poisson brackets of free fields. In light of this observation, and given its close analogy with the spin-two analysis developed in Subsections~\ref{subsec. AdS-diag} and~\ref{subsec. flat-boson}, this construction will serve as our starting point for investigating the $\Lambda \to 0$ case.

Following the same type of strategy, we introduce four celestial bosons encoding the boundary degrees of freedom,
\begin{equation}
    \mathscr{P}_{(L,W)} = \psi_{(L,W)}' \, , \qquad 
    \bar{\mathscr{P}}_{(L,W)} = \bar{\psi}_{(L,W)}' + u \, \psi_{(L,W)}'' \, ,
\end{equation}
which are defined as the Carrollian descendants of their Lorentzian avatars~\eqref{eq. s=3 bosons}:
\begin{subequations} \label{eq. flat s=3 flat prescriptions}
    \begin{align}
        &\mathscr{P}_{(L,W)} = \lim_{\ell \to \infty} \frac{1}{2} \left( \mathscr{J}_{(L,W)} - \bar{\mathscr{J}}_{(L,W)} \right) ,\\
        &\bar{\mathscr{P}}_{(L,W)} = \lim_{\ell \to \infty} \frac{\ell}{2} \left( \mathscr{J}_{(L,W)} + \bar{\mathscr{J}}_{(L,W)} \right) .
    \end{align}
\end{subequations}
In the flat limit, this leads to the higher-spin version of the Carrollian free field gauge \eqref{eq. flat bosonic gauge}, defined on-shell by the asymptotic solution space
\begin{subequations} \label{eq. hs boson gauge}
    \begin{align}
        \begin{split}
        \alpha_\phi &= J_1 + \psi_{(L)}'(\phi) \, J_0 + \left( \bar{\psi}_{(L)}'(\phi) + u \, \psi_{(L)}''(\phi) \right) P_0\\
        & \quad + \psi_{(W)}'(\phi) \, U_0 + \left( \bar{\psi}_{(W)}'(\phi) + u \, \psi_{(W)}''(\phi) \right) V_0 \, ,
        \end{split}\\
        \alpha_u &= P_1 + \psi_{(L)}'(\phi) \, P_0 + \psi_{(W)}'(\phi) \, V_0 \, .
    \end{align}
\end{subequations}
Note again the distinction with the boundary conditions of~\cite{Ammon:2017vwt}, where the asymptotic fall-offs~\eqref{eq. flat bc} were not imposed, while chemical potentials were introduced in the temporal component $\alpha_u$. We now demonstrate that the gauge \eqref{eq. hs boson gauge} yields a holographic realization of the spin-three extension of the $\mathfrak{bms}_3$ charge algebra in terms of Carrollian free fields.

\subsection{Asymptotic symmetries} \label{subsec. hs-asympt}

The solution space~\eqref{eq. hs boson gauge} admits the following residual symmetries:
\begin{equation}
    \Lambda(u,r,\phi) = b^{-1}(r) \left( \sum_{i=-1}^1 \left( \epsilon^i_{(L)} J_i + \sigma^i_{(L)} P_i \right) + \sum_{n=-2}^2 \left( \epsilon^n_{(W)} U_n + \sigma^n_{(W)} V_n \right) \right) b(r) \, ,
\end{equation}
where the asymptotic symmetry generators are given by
\begin{subequations}
    \begin{align}
        \epsilon^0_{(L,W)} &= \gamma_{(L,W)}(\phi) \, ,\\
        \sigma^0_{(L,W)} &= \bar{\gamma}_{(L,W)}(\phi) + u \, \gamma_{(L,W)}'(\phi) \, ,
    \end{align}
\end{subequations}
and
\begin{subequations}
    \begin{align}
        &\epsilon^1_{(L)} \pm \epsilon^1_{(W)}
        = \text{e}^{\psi_{(L)} \pm 2\psi_{(W)}}\left(
        \kappa_{4,5}
        - \int_{\phi_0}^\phi \mathrm{d}\phi' \,
        \text{e}^{-\psi_{(L)}(\phi') \mp 2\psi_{(W)}(\phi')}
        (\gamma_{(L)}(\phi') \pm \gamma_{(W)}(\phi'))
        \right) ,
        \\
        &\sigma^1_{(L)} \pm \sigma^1_{(W)}
        = u\left(-\gamma_{(L)} \mp 2\gamma_{(W)}\right)
        \notag\\
        &\quad + \text{e}^{\psi_{(L)} \pm 2\psi_{(W)}}\Bigg(
        \kappa_{10,11}
        - \int_{\phi_0}^\phi \mathrm{d}\phi'\, \Big(
        \text{e}^{-\psi_{(L)}(\phi') \mp 2\psi_{(W)}(\phi')}
        \big(\bar{\gamma}_{(L)}(\phi') + u\gamma'_{(L)}(\phi')
        \notag\\
        &\qquad\qquad \pm 2(\bar{\gamma}_{(W)}(\phi') + u\gamma'_{(W)}(\phi'))\big)
        \notag\\
        &\qquad\qquad + \Big(-\kappa_{4,5}
        + \int_{\phi_0'}^{\phi'} \mathrm{d}\phi'' \,
        \text{e}^{-\psi_{(L)}(\phi'') \mp 2\psi_{(W)}(\phi'')}
        (\gamma_{(L)}(\phi'') \pm 2\gamma_{(W)}(\phi''))\Big)
        \notag\\
        &\qquad\qquad \times (\bar{\psi}_{(L)}'(\phi') \pm 2\bar{\psi}_{(W)}'(\phi'))
        \Big)
        \notag\\
        &\quad + u \big(\psi'_{(L)} \pm 2\psi'_{(W)}\big)
        \Big(
        \kappa_{4,5}
        - \int_{\phi_0}^\phi \mathrm{d}\phi' \,
        \text{e}^{-\psi_{(L)}(\phi') \mp 2\psi_{(W)}(\phi')}
        (\gamma_{(L)}(\phi') \pm 2\gamma_{(W)}(\phi'))
        \Big)
        \Bigg) \, .
    \end{align}
\end{subequations}
The pure gauge generators take the form
\begin{subequations} \label{eq. hs pure gauge resid}
    \begin{align}
        \epsilon^{-2}_{(W)} &= \kappa_1 \, \mathrm{e}^{-2 \psi_{(L)}} \, ,\\
        \sigma^{-2}_{(W)} &= \mathrm{e}^{-2\psi_{(L)}}\left(\kappa_7 - 2\kappa_1\left(\bar{\psi}_{(L)} + u\psi'_{(L)}\right)\right) \, ,\\
        \epsilon^{-1}_{(L)} \pm \epsilon^{-1}_{(W)} 
        &= \mathrm{e}^{-\psi_{(L)} \mp 2\psi_{(W)}} 
        \left(
        \kappa_{2,3} 
        \mp 4\kappa_1 \int_{\phi_0}^\phi \mathrm{d}\phi' \, 
        \mathrm{e}^{-\psi_{(L)}(\phi') \mp 2\psi_{(W)}(\phi')}
        \right) ,\\
        \begin{split}
        \sigma^{-1}_{(L)} \pm \sigma^{-1}_{(W)} 
        &= \kappa_{8,9}\,\mathrm{e}^{-\psi_{(L)} \mp 2\psi_{(W)}}
        \mp 4u\kappa_1\,\mathrm{e}^{-2\psi_{(L)}} \\
        &\quad + \mathrm{e}^{-\psi_{(L)} \mp 2\psi_{(W)}}
        \int_{\phi_0}^\phi \mathrm{d}\phi' \, \Big(
        \mp 4\,\mathrm{e}^{-\psi_{(L)}(\phi') \pm 2\psi_{(W)}(\phi')}
        \big(\kappa_7 - 2\kappa_1 \bar{\psi}_{(L)}(\phi')\big) \\
        &\qquad - \Big(
        \kappa_{2,3} 
        \mp 4\kappa_1 \int_{\phi_0'}^{\phi'} \mathrm{d}\phi'' \,
        \mathrm{e}^{-\psi_{(L)}(\phi'') \pm 2\psi_{(W)}(\phi'')}
        \Big)
        \big(\bar{\psi}'_{(L)}(\phi') \pm 2\bar{\psi}'_{(W)}(\phi')\big)
        \Big) \\
        &\quad - u\,\mathrm{e}^{-\psi_{(L)} \mp 2\psi_{(W)}}
        \Big(
        \kappa_{2,3}
        \mp 4\kappa_1 \int_{\phi_0}^\phi \mathrm{d}\phi' \,
        \mathrm{e}^{-\psi_{(L)}(\phi') \pm 2\psi_{(W)}(\phi')}
        \Big)
        \big(\psi'_{(L)} \pm 2\psi'_{(W)}\big) \, .
        \end{split}
    \end{align}
\end{subequations}
Notice that in the last equations, we have deliberately omitted the modes $\epsilon_{(W)}^{(2)}$ and $\sigma_{(W)}^{(2)}$, since their explicit expressions are cumbersome and not useful for the subsequent analysis.\footnote{Actually, it is still interesting to note that they depend on the constants $\kappa_{4,5,6}$ and $\kappa_{4,5,6,10,11,12}$, respectively, which, as we will show explicitly below, do not affect either the asymptotic charges or the pure gauge charges. These should therefore be interpreted as stabilizers of the residual symmetries.}

We also emphasize that we have already distinguished between asymptotic parameters $\gamma_{(L,W)}(\phi)$ and $\bar{\gamma}_{(L,W)}(\phi)$, and pure gauge parameters $\kappa_{1,2,3,7,8,9}$ (constants). This distinction will be further justified at the level of asymptotic surface charges and Carrollian screening charges, but it is already manifest in the gauge transformations of the physical fields:
\begin{equation}
    \delta_\Lambda \left( \psi_{(L)}' \pm \frac{2}{3} \psi_{(W)}' \right) 
    = \gamma_{(L)}' \pm \frac{2}{3} \gamma_{(W)}' 
    + 2 \, \mathrm{e}^{- \psi_{(L)} \mp 2 \psi_{(W)}} 
    \left( \kappa_{2,3} \mp 4 \kappa_1 \int^\phi_{\phi_0} \mathrm{d}\phi ' \, \mathrm{e}^{-\psi_{(L)} \pm 2 \psi_{(W)}} \right) ,
\end{equation}
and
\begin{equation}
    \begin{split}
    \delta_\Lambda \left( \bar{\psi}_{(L)}' \pm \frac{2}{3} \bar{\psi}_{(W)}' \right) 
    &= \bar{\gamma}_{(L)}' \pm \frac{2}{3} \bar{\gamma}_{(W)}' 
    + 2 \, \mathrm{e}^{- \psi_{(L)} \mp 2 \psi_{(W)}} \Bigg\{ \kappa_{8,9}\\
    & \quad \mp \int^\phi_{\phi_0} \mathrm{d}\phi ' \Bigg[ 4 \, \mathrm{e}^{-\psi_{(L)} \pm 2 \psi_{(W)}} \left( \kappa_7 - 2 \kappa_1 \bar{\psi}_{(L)} \right)\\
    &\quad \pm \left( \kappa_{2,3} \mp 4 \kappa_1 \int^{\phi '}_{\phi_0'} \mathrm{d} \phi '' \, \mathrm{e}^{-\psi_{(L)} \pm 2 \psi_{(W)}} \right) 
    \left( \bar{\psi}_{(L)}' \pm 2 \bar{\psi}_{(W)}' \right) \Bigg] \Bigg\} \, .
    \end{split}
\end{equation}
Under the modified bracket~\eqref{eq. modifiedbracket}, the residual symmetry algebra is Abelian:
\begin{equation}
    \gamma^{(12)}_{(L,W)} = 0 \, , \qquad \bar{\gamma}^{(12)}_{(L,W)} = 0 \, .
\end{equation}
Note again that~\cite{Ammon:2017vwt} restricts attention to the parameters associated with the zero modes of the Poincar\'e algebra, leaving the role of pure gauge symmetries unexplored.

Then, the asymptotic corner charges take the form
\begin{equation}
    H_\Lambda = - \frac{1}{8 \pi G} \int \mathrm{d}\phi \left[ \gamma_{(L)} \bar{\psi}_{(L)}' + \bar{\gamma}_{(L)} \psi_{(L)}' + \frac{4}{3} \left( \gamma_{(W)} \bar{\psi}_{(W)}' + \bar{\gamma}_{(W)} \psi_{(W)}' \right) \right] ,
\end{equation}
and indeed depend solely on the residual symmetry functions defined on the celestial circle. Under the Poisson structure \eqref{eq. PB}, these charges furnish a projective representation of this Abelian algebra, yielding a higher-spin Carrollian Heisenberg algebra:
\begin{equation} \label{eq. hs charge algebra}
    \mathrm{i} \{\Psi_{(L)}^n,\bar{\Psi}_{(L)}^m\} = - \frac{c_2}{12} n \, \delta_{n+m,0} \, , \qquad 
    \mathrm{i} \{\Psi_{(W)}^n,\bar{\Psi}_{(W)}^m\} = - \frac{c_2}{9} n \, \delta_{n+m,0} \, ,
\end{equation}
with all other brackets vanishing, and with BMS central charge~\eqref{eq. flat central charge}. Here, we have introduced the Fourier modes as
\begin{subequations}
    \begin{align}
        &\Psi_{(L)}^n := H_\Lambda(\gamma_{(L,W)} = 0,\, \bar{\gamma}_{(L)} \sim \mathrm{e}^{\mathrm{i}n\phi},\, \bar{\gamma}_{(W)} = 0) \, ,\\
        &\bar{\Psi}_{(L)}^n := H_\Lambda(\gamma_{(L)} \sim \mathrm{e}^{\mathrm{i}n\phi},\, \gamma_{(W)} = 0,\, \bar{\gamma}_{(L,W)} = 0) \, ,\\
        &\Psi_{(W)}^n := H_\Lambda(\gamma_{(L,W)} = 0,\, \bar{\gamma}_{(L)} = 0,\, \bar{\gamma}_{(W)} \sim \mathrm{e}^{\mathrm{i}n\phi}) \, ,\\
        &\bar{\Psi}_{(W)}^n := H_\Lambda(\gamma_{(L)} = 0,\, \gamma_{(W)} \sim \mathrm{e}^{\mathrm{i}n\phi},\, \bar{\gamma}_{(L,W)} = 0) \, .
    \end{align}
\end{subequations}
This provides a holographic realization of the asymptotic symmetries of higher-spin fields coupled to asymptotically flat spacetimes in terms of a reduced symplectic structure of Carrollian free fields.

We now conclude this Section by demonstrating that the boundary symplectic form admits degenerate directions which, upon reduction, give rise to the physical observables associated with the higher-spin Bondi aspects.

\subsection{Carrollian screening charges} \label{subsec. hs-screening}

From the Poisson structure~\eqref{eq. hs charge algebra}, we in particular deduce that
\begin{subequations}
    \begin{align}
        &\{\psi_{(L)}'(\phi),\bar{\psi}_{(L)}(\phi')\} = \{\bar{\psi}_{(L)}'(\phi),\psi_{(L)}(\phi')\} = 8 \pi G \, \delta(\phi - \phi') \, ,\\
        &\{\psi_{(W)}'(\phi),\bar{\psi}_{(W)}(\phi')\} = \{\bar{\psi}_{(W)}'(\phi),\psi_{(W)}(\phi')\} = 6 \pi G \, \delta(\phi - \phi') \, ,
    \end{align}
\end{subequations}
and, by the chain rule,
\begin{subequations}
    \begin{align}
        &\{\bar{\psi}_{(L)}'(\phi),\mathrm{e}^{-\psi_{(L)}(\phi')}\} = - \mathrm{e}^{-\psi_{(L)}(\phi')} \{\psi_{(L)}'(\phi),\psi_{(L)}(\phi')\} \, ,\\
        &\{\bar{\psi}_{(W)}'(\phi),\mathrm{e}^{-\psi_{(W)}(\phi')}\} = - \mathrm{e}^{-\psi_{(W)}(\phi')} \{\psi_{(W)}'(\phi),\psi_{(W)}(\phi')\} \, .
    \end{align}
\end{subequations}
It then follows that there exist six functionals generating the residual pure gauge transformations, acting on the physical fields as
\begin{equation}
    \{S_{\Lambda},\psi_{(L,W)}'\} = \delta_{\Lambda} \psi_{(L,W)}' \, , 
    \qquad 
    \{S_{\Lambda},\bar{\psi}_{(L,W)}'\} = \delta_{\Lambda} \bar{\psi}_{(L,W)}' \, ,
\end{equation}
where $\Lambda$ is restricted to the pure gauge sector~\eqref{eq. hs pure gauge resid}, i.e. non-vanishing only for the constants $\kappa_{1,2,3,7,8,9}$.\footnote{The remaining constants again stabilize the space of residual symmetries.}

These can be organized into four independent Carrollian screening charges:
\begin{subequations} \label{eq. hs screening 1}
    \begin{align}
        S_{\kappa_8} &= \frac{\kappa_8}{8 \pi G} \int \mathrm{d}\phi \, \mathrm{e}^{-\psi_{(L)} - 2 \psi_{(W)}} \, ,\\
        S_{\kappa_9} &= \frac{\kappa_9}{8 \pi G} \int \mathrm{d}\phi \, \mathrm{e}^{-\psi_{(L)} + 2 \psi_{(W)}} \, ,\\
        S_{\kappa_2} &= - \frac{\kappa_2}{8 \pi G} \int \mathrm{d}\phi \left( \bar{\psi}_{(L)} + 2 \bar{\psi}_{(W)} \right) \mathrm{e}^{-\psi_{(L)} - 2 \psi_{(W)}} \, ,\\
        S_{\kappa_3} &= - \frac{\kappa_3}{8 \pi G} \int \mathrm{d}\phi \left( \bar{\psi}_{(L)} - 2 \bar{\psi}_{(W)} \right) \mathrm{e}^{-\psi_{(L)} + 2 \psi_{(W)}} \, ,
    \end{align}
\end{subequations}
together with two composite Carrollian screening charges, which can be obtained from Poisson brackets of the previous ones:
\begin{subequations} \label{eq. hs screening 2}
    \begin{align}
        \begin{split}
        S_{\kappa_7} &= - \frac{\kappa_7}{2 \pi G} \int \mathrm{d} \phi \left[ \mathrm{e}^{-\psi_{(L)} - 2 \psi_{(W)}} \int^\phi_{\phi_0} \mathrm{d}\phi' \, \mathrm{e}^{-\psi_{(L)} + 2 \psi_{(W)}} \right.\\
        &\qquad\qquad\left. - \mathrm{e}^{-\psi_{(L)} + 2 \psi_{(W)}} \int^\phi_{\phi_0} \mathrm{d}\phi' \, \mathrm{e}^{-\psi_{(L)} - 2 \psi_{(W)}} \right] ,
        \end{split}\\
        \begin{split}
        S_{\kappa_1} &= - \frac{\kappa_1}{2 \pi G} \int \mathrm{d} \phi \Bigg[ \mathrm{e}^{-\psi_{(L)} - 2 \psi_{(W)}} \left( \bar{\psi}_{(L)} + 2 \bar{\psi}_{(W)} \right) \int^\phi_{\phi_0} \mathrm{d}\phi' \, \mathrm{e}^{-\psi_{(L)} + 2 \psi_{(W)}}\\
        &\qquad\qquad - \mathrm{e}^{-\psi_{(L)} + 2 \psi_{(W)}} \left( \bar{\psi}_{(L)} - 2 \bar{\psi}_{(W)} \right) \int^\phi_{\phi_0} \mathrm{d}\phi' \, \mathrm{e}^{-\psi_{(L)} - 2 \psi_{(W)}} \Bigg] .
        \end{split}
    \end{align}
\end{subequations}
This is equivalent to the functional integration of $q_{\delta \Lambda}$ within the covariant phase space formalism~\eqref{eq. generators pure gauge}, as there exists a distinct splitting between $q_{\delta \Lambda}$ and the Iyer--Wald charge~$k_\Lambda$ in terms of the residual parameters entering the Noether charge~$q_\Lambda$.

The realization of the asymptotic symmetry algebra in terms of free fields, together with the presence of screening charges, indicates that the putative dual field theory takes the form of a Carrollian analogue of the higher-spin Coulomb gas. Such a formulation is particularly well suited for quantization, as emphasized in~\cite{BALOG199076,Campoleoni:2017xyl,Fredenhagen:2025aqd}, and thus constitutes a promising step toward a concrete realization of higher-spin holography in flat space.

\subsection{Carrollian Miura transformations} \label{subsec. hs-Miura}

Since the boundary symplectic structure possesses degenerate directions, it can be reduced by quotienting out the pure gauge orbits, thereby allowing us to identify the genuine physical observables. The Carrollian screening charges~\eqref{eq. hs screening 1}--\eqref{eq. hs screening 2} provide a convenient way to determine them by imposing the condition~\eqref{eq. screening condition Miura}. We then find that the physical observables are given by the following combinations of the Carrollian higher-spin free fields:
\begin{subequations} \label{eq. hs Miura}
    \begin{align}
        M_{(L)} &= \frac{1}{12} \left[ - 3 \left( \psi_{(L)}' \right)^2 - 4 \left( \psi_{(W)}' \right)^2 - 6 \psi_{(L)}'' \right] ,\\
        L_{(L)} &= \frac{1}{6} \left[ - 3 \psi_{(L)}' \bar{\psi}_{(L)}' - 4 \psi_{(W)}' \bar{\psi}_{(W)}' - 3 \bar{\psi}_{(L)}'' \right] ,\\
        \begin{split}
        M_{(W)} &= \frac{1}{108} \bigg[ \psi_{(W)}' \left( 18 \left( \psi_{(L)}' \right)^2 - 8 \left( \psi_{(W)}' \right)^2 + 9 \left( \psi_{(L)}'' \right)^2 \right)\\
        &\quad + 27 \psi_{(L)}' \psi_{(W)}'' + 9 \psi_{(W)}''' \bigg] \, ,
        \end{split}\\
        \begin{split}
        L_{(W)} &= \frac{1}{36} \bigg[ \bar{\psi}_{(W)}' \left( 6 \left( \psi_{(L)}' \right)^2 - 8 \left( \psi_{(W)}' \right)^2 + 3 \psi_{(L)}'' \right) + 3 \psi_{(W)}' \bar{\psi}_{(L)}''\\
        &\quad + 9 \psi_{(L)}' \bar{\psi}_{(W)}'' + 3 \bar{\psi}_{(L)}' \left( 4 \psi_{(L)}' \psi_{(W)}' + 3 \psi_{(W)}'' \right) + 3 \bar{\psi}_{(W)}''' \bigg] \, ,
        \end{split}
    \end{align}
\end{subequations}
such that, indeed, for $\Lambda$ restricted to the pure gauge sector~\eqref{eq. hs pure gauge resid},
\begin{equation}
    \delta_{\Lambda} M_{(L,W)} = 0 \, , \qquad \delta_{\Lambda} L_{(L,W)} = 0 \, .
\end{equation}
These combinations correspond precisely to the higher-spin Carrollian Miura transformations, originally identified from a purely field-theoretic perspective in~\cite{Fredenhagen:2025aqd}, and now realized holographically through the coupling of higher-spin fields to asymptotically flat spacetimes, where they also admit a clear physical interpretation. Note again their distinct cosmological mention in \cite{Ammon:2017vwt}.

From the viewpoint of the Chern--Simons formulation, this realization arises from a change of gauge from the higher-spin Bondi gauge~\eqref{eq. hs Bondi gauge} to the higher-spin Carrollian bosonic gauge~\eqref{eq. hs boson gauge}. In particular, the two gauges are related by a gauge transformation of the form
\begin{equation}
    \alpha_\mathrm{diag} = b^{-1} \left( \alpha_\mathrm{BMS} + \mathrm{d} \right) b \, ,
\end{equation}
with
\begin{equation}
    \begin{split}
    b &= \exp \bigg[ \frac{1}{2} \psi_{(L)}' J_{-1} + \frac{1}{2} \left( \bar{\psi}_{(L)}' + u \, \psi_{(L)}'' \right) P_{-1} + \frac{1}{3} \psi_{(W)}' U_{-1}\\
    &\quad + \frac{1}{3} \left( \bar{\psi}_{(W)}' + u \, \psi_{(W)}'' \right) V_{-1} - \frac{1}{12} \left( 2 \psi_{(L)}' \psi_{(W)}' + \psi_{(W)}'' \right) U_{-2}\\
    &\quad - \frac{1}{12} \bigg( \bar{\psi}_{(W)}'' + 2 \psi_{(W)}' \left( \bar{\psi}_{(L)}' + u \, \psi_{(L)}'' \right) + u \, \psi_{(W)}'''\\
    &\quad + 2 \psi_{(L)}' \left( \bar{\psi}_{(W)}' + u \, \psi_{(W)}'' \right) \bigg) V_{-2} \bigg] \, ,
    \end{split}
\end{equation}
provided that the higher-spin Bondi mass and angular momentum aspects can be expressed in terms of the celestial bosons through the higher-spin Carrollian Miura transformations~\eqref{eq. hs Miura}.


\section{Going beyond spin three} \label{sec. beyond}

In this final Section, we generalize the previous constructions to gauge algebras $\mathfrak{isl}(s)$ for arbitrary $s$, focusing on algebraic aspects rather than bosonic ones, as these are better suited to the formality and complexity of such a generalization. One can realize these algebras as traceless matrices over the dual numbers,
\begin{equation}
    \mathfrak{isl}(s) \simeq \mathfrak{sl}(s) \otimes \mathbb{R}[\theta] \simeq \{ A_0+\theta A_1\,|\, A_0,A_1 \in \mathfrak{sl}(s) \}\,,
\end{equation}
where $\theta$ is a nilpotent parameter satisfying $\theta^2=0$. For instance, for $\mathfrak{isl}(2)$, the generators admit a matrix representation given by
\begin{align}
    J_1 &= \begin{pmatrix}
        0 & 0 \\ -1 & 0
    \end{pmatrix} , &
    J_0 &= \frac{1}{2} \begin{pmatrix}
        1 & 0 \\ 0 & -1
    \end{pmatrix} , &
    J_{-1} &= \begin{pmatrix}
        0 & 1 \\ 0 & 0 
    \end{pmatrix} , &
    P_i &= \theta\, J_i \, .
\end{align}
An invariant bilinear form is then defined by 
\begin{equation}
    \mathrm{Tr}(A B) = \alpha_s \,\mathrm{tr}(A B) \, ,
\end{equation}
where $\mathrm{tr}(\bullet)$ denotes the matrix trace. The normalization constant is chosen to be
\begin{equation}
    \alpha_s=\frac{12}{s(s^2-1)}\,.
\end{equation}
Concretely, this yields
\begin{equation}
    \mathrm{Tr}\big((A_0+\theta A_1)(B_0+\theta B_1)\big) 
    = \alpha_s \,\mathrm{tr}(A_0 B_1 + A_1 B_0)\, .
\end{equation}
Within $\mathfrak{isl}(s)$, one can identify $\mathfrak{sl}(2)$ as a subalgebra by representing its generators as
\begin{align}
    J_1 &= -\sum_{j=1}^{s-1} e_{j+1,j} \, , &
    J_0 &= \sum_{j=1}^s \left(\frac{s+1}{2}-j\right) e_{jj} \, , &
    J_{-1} &= \sum_{j=1}^{s-1} j \left(s-j\right) e_{j,j+1} \, .
\end{align}
Here, the $e_{ij}$ denote the elementary matrices with a single non-vanishing entry equal to~$1$ in row $i$ and column $j$. With respect to the adjoint action of this subalgebra, one can decompose $\mathfrak{isl}(s)$ into irreducible representations of $\mathfrak{sl}(2)$. Both the $\mathfrak{sl}(s)$ and $\theta \mathfrak{sl}(s)$ sectors decompose into irreducible representations of $\mathfrak{sl}(2)$ of spin $\ell=1,\dots,s-1$, with generators
\begin{align}
    U^{(\ell+1)}_n \, , \qquad 
    V^{(\ell+1)}_n = \theta \, U^{(\ell+1)}_n \, ,
    \qquad \ell=1,\dots,s-1 \, , \qquad n=-\ell,\dots,\ell \, .
\end{align}
They satisfy
\begin{equation}
    \big[J_i , U^{(\ell+1)}_n\big] = (\ell\, i - n)\, U^{(\ell+1)}_{i+n} \,, \qquad
    \big[J_i , V^{(\ell+1)}_n\big] = (\ell\, i - n)\, V^{(\ell+1)}_{i+n} \, ,
\end{equation}
where, in this equation, $-1 \leq i \leq 1$. In this realization, the generators with positive mode numbers are strictly lower triangular matrices, the generators with negative mode numbers are strictly upper triangular matrices, and the generators $U^{\ell+1}_0$ and $V^{\ell+1}_0$ are diagonal matrices. We thus gradually recover the idea of a diagonal gauge.

\subsection{Flat diagonal gauge}

Within the algebraic framework described above, one can generalize the Carrollian free-field gauge defined for spin two in~\eqref{eq. flat bosonic gauge} and spin three in~\eqref{eq. hs boson gauge} to arbitrary higher spins $s$, as an $\mathfrak{isl}(s)$ diagonal gauge obtained by setting
\begin{subequations} \label{eq. hs diag gauge}
    \begin{align}
        \alpha_\phi &= J_1 + \left(1 + \theta \, u \, \partial_\phi \right) D(\phi) \,, \\
        \alpha_u &= \theta\,\alpha_\phi = P_1 + \theta\, D(\phi)\,,
    \end{align}
\end{subequations}
with a diagonal matrix $D(\phi)=D_0(\phi)+\theta\,D_1(\phi)$, or, equivalently,
\begin{subequations}
    \begin{align}
        \alpha_\phi &= J_1 + D_0 + \theta \left( D_1 + u \, D_0 ' \right) ,\\
        \alpha_u &= P_1 + \theta \, D_0 \, .
    \end{align}
\end{subequations}
This indeed matches the definitions~\eqref{eq. flat bosonic gauge} and~\eqref{eq. hs boson gauge} when reduced to $s=2$ and $s=3$, respectively. We continue to assume that we are working in the radial gauge~\eqref{eq. radialgauge}. These components~\eqref{eq. hs diag gauge} solve the equations of motion~\eqref{eq. eomCS}, as the CS connection is readily seen to satisfy
\begin{equation}
    \partial_u \alpha_\phi = \partial_\phi \alpha_u = \theta \,\partial_\phi \alpha_\phi \, .
\end{equation}
Note that one can combine the expressions of the asymptotic solution space into
\begin{equation}
    \alpha= \alpha_\phi \mathrm{d}\phi + \alpha_u\mathrm{d}u = \big(J_1 + (1+\theta u \partial_\phi)D(\phi)\big)\big(\mathrm{d}\phi + \theta \mathrm{d}u\big)\,.
\end{equation}

A gauge transformation of the form~\eqref{eq. InfGaugeSym}
\begin{equation}
    \delta_\Lambda \alpha = \mathrm{d}\Lambda + [\alpha,\Lambda]
\end{equation}
respects in particular the condition $\alpha_u=\theta \, \alpha_\phi$ provided that $\partial_u \Lambda = \theta \,\partial_\phi \Lambda$, in which case $\Lambda$ takes the form
\begin{equation}
    \Lambda(u,\phi)= \Lambda_0(\phi) + \theta \big(u\,\Lambda_0'(\phi) + \Lambda_1(\phi)\big) \, .
\end{equation}
Such residual symmetries automatically ensure $\partial_u \delta_\Lambda\alpha_\phi = \theta\,\partial_\phi \delta_\Lambda\alpha_\phi$, and thus the remaining condition on $\Lambda$ reads
\begin{equation}
    \delta_\Lambda D= \big(\partial_\phi\Lambda + [\alpha_\phi,\Lambda]\big)\big|_{u=0} \, , 
\end{equation}
which implies
\begin{subequations}
    \begin{align}
        \label{deltaD0}
        \delta_\Lambda D_0 &= \Lambda_0' + [J_1+D_0,\Lambda_0] \, ,\\
        \label{deltaD1}    
        \delta_\Lambda D_1 &= \Lambda_1' + [D_1,\Lambda_0] + [J_1+D_0,\Lambda_1] \, .
    \end{align}
\end{subequations}

\subsection{Asymptotic symmetries}

To analyze the residual symmetries further, we introduce some notation for matrices. We decompose a generic matrix $M$ into its strictly upper triangular part $M^>$, diagonal part~$M^\mathrm{d}$, and strictly lower triangular part $M^<$,
\begin{equation}
    M = M^> + M^\mathrm{d} + M^< \, .
\end{equation}
The upper and lower triangular parts are then defined as
\begin{equation}
    M^\geq = M^> + M^\mathrm{d} \, , \qquad
    M^\leq = M^\mathrm{d} + M^< \, ,
\end{equation}
respectively. Since $J_1$ has non-vanishing entries only on the first lower sub-diagonal, the gauge transformations~\eqref{deltaD0} and~\eqref{deltaD1} can be split into two parts. One involves only the strictly upper triangular gauge parameters $\Lambda_0^>$ and $\Lambda_1^>$,
\begin{subequations}
    \begin{align}
        \label{lambda0 sut}
        0 &= (\Lambda_0^>)' + [J_1,\Lambda_0^>]^> + [D_0,\Lambda_0^>] \, ,\\
        \label{lambda1 sut}
        0 &= (\Lambda_1^>)' + [D_1,\Lambda_0^>] + [J_1,\Lambda_1^>]^> + [D_0,\Lambda_1^>] \, ,\\
        \label{gaugetrans D0}
        \delta_{\Lambda^>} D_0 &= [J_1,\Lambda_0^>]^\mathrm{d} \, ,\\
        \label{gaugetrans D1}
        \delta_{\Lambda^>} D_1 &= [J_1,\Lambda_1^>]^\mathrm{d} \, .
    \end{align}
\end{subequations}
In comparison with the cases $s=2$ in~\eqref{eq. flat bosonic resid} and $s=3$ in~\eqref{eq. hs pure gauge resid}, we interpret these residual transformations as pure gauge. This will be confirmed explicitly in the next Subsection \ref{subsec. beyond screening} through the computation of the associated screening charges. The linear system~\eqref{lambda0 sut}--\eqref{lambda1 sut} can be solved iteratively, starting from the top-right entries of $\Lambda^>$. Its solutions are parametrized by $s(s-1)$ integration constants, which in turn generate the pure gauge transformations~\eqref{gaugetrans D0}--\eqref{gaugetrans D1} of $D_0$ and $D_1$. These correspond precisely to a subset of the constants $\{\kappa_n\}$ (for $n = 1, \dots, 2s(s-1)$) introduced in the previous Sections.

A second type of gauge transformations involves the lower triangular part $\Lambda^\leq$ and corresponds to the asymptotic symmetries. This is again consistent with the previous analyses and will also be confirmed at the level of the surface charges. For these symmetries, the equations read
\begin{subequations}
    \begin{align}
        \label{asympttrans D0}
        \delta_{\Lambda^\leq} D_0 &= (\Lambda_0^\mathrm{d})'  \, ,\\
        \label{asympttrans D1}
        \delta_{\Lambda^\leq} D_1 &= (\Lambda_1^\mathrm{d})' \, ,\\
        \label{lambda0 lt}
        0 & = (\Lambda_0^<)' + [J_1,\Lambda_0^\leq] + [D_0,\Lambda_0^<] \, ,\\
        \label{lambda1 lt}
        0 &= (\Lambda_1^<)' + [D_1,\Lambda_0^<] + [J_1,\Lambda_1^\leq] + [D_0,\Lambda_1^<] \, .
    \end{align}
\end{subequations}
From~\eqref{asympttrans D0} and~\eqref{asympttrans D1}, we see that the transformation of $D$ under the asymptotic symmetries is entirely determined by the diagonal part $\Lambda^\mathrm{d}$ of the gauge parameter. The strictly lower triangular part $\Lambda^<$ is then determined by the system of differential equations~\eqref{lambda0 lt} and~\eqref{lambda1 lt} up to integration constants, which do not affect the transformation $\delta_{\Lambda^\leq}D$.

This leads to the Iyer--Wald density~\eqref{eq. kCS}
\begin{equation}
    k_\Lambda = - 2 \kappa \, \mathrm{Tr} \! \left( \Lambda^\mathrm{d} \, \delta D \right) \mathrm{d}\phi \, .
\end{equation}
Assuming $\delta \Lambda^\mathrm{d} = 0$, as already done for spin two and three, this density can be integrated to yield the following surface charge: 
\begin{equation} \label{eq. HLambda hs}
    H_{\Lambda} = -2 \kappa \int \mathrm{d}\phi\,\mathrm{Tr} \! \left(\Lambda^\mathrm{d}\,D\right)
    = -2 \kappa\,\alpha_s\int \mathrm{d}\phi \,\mathrm{tr} \! \left(\Lambda_1^\mathrm{d} \,D_0 + \Lambda_0^\mathrm{d}\,D_1\right) .
\end{equation}
We observe that only the diagonal components $\Lambda^\mathrm{d}$ contribute to the Hamiltonian above, thereby confirming the nomenclature introduced earlier for the residual symmetries. The asymptotic charges then imply the following Poisson brackets (via~\eqref{eq. PB}):
\begin{subequations} \label{eq. PB hs}
    \begin{align}
        \{ (D_0)_{ii}(\phi_1) ,(D_0)_{jj}(\phi_2)\} & = 0 \, , \\
        \{ (D_1)_{ii}(\phi_1) ,(D_1)_{jj}(\phi_2)\} &= 0 \, , \\
        \{ (D_0)_{ii}(\phi_1),(D_1)_{jj}(\phi_2) \} &= -\frac{1}{2\kappa\,\alpha_s} \left(\delta_{ij}-\frac{1}{s}\right)\,\delta'(\phi_1-\phi_2) \, .
\end{align}
\end{subequations}
We thus obtain a Carrollian free-field algebra for any value $s$ of the higher-spin coupling. One can show that this corresponds to a projective representation of the residual symmetry algebra, as in~\eqref{eq. chargalg}. Indeed, since we have assumed that the asymptotic components of the residual gauge parameter are field-independent, the modified bracket~\eqref{eq. modifiedbracket} reduces to the ordinary commutator for these components. In addition, the physical asymptotic symmetries are actually parametrized by the diagonal part of the residual parameter. Because diagonal matrices commute, this sector is therefore Abelian:
\begin{equation}
    \Lambda^\mathrm{d}_{(12)} := [\Lambda^\mathrm{d}_{(1)},\Lambda^\mathrm{d}_{(2)}] = 0 \, ,
\end{equation}
and the above charge algebra~\eqref{eq. PB hs} accordingly provides its central extension.

\subsection{Carrollian screening charges} \label{subsec. beyond screening}

We now discuss the pure gauge symmetries in more details, and their screening charges. It turns out that, under the assumption $\delta \Lambda^\mathrm{d} = 0$, the diagonal components of the residual symmetries cannot contribute to $q_{\delta \Lambda}$ in the decomposition~\eqref{eq. kqqdelta}, but only to $k_\Lambda$, as seen above in \eqref{eq. HLambda hs}. We thus recover once again a clear splitting between these two quantities, here in a more geometric way. This shows that the underlying reason for this splitting lies in the field-independence of the full asymptotic component in the mode decomposition of the gauge parameter, which we have indeed assumed. Actually, we have seen for $s=2$ and $s=3$ that $\delta \epsilon^{0}_{(L,W)} = 0$ and $\delta \sigma^{0}_{(L,W)} = 0$. This implies that we can again follow the CPS steps and functionally integrate $q_{\delta \Lambda}$ to obtain the associated Carrollian screening charges~\eqref{eq. generators pure gauge}. From the general expression~\eqref{eq. qLambda}
\begin{equation}
    q_{\delta \Lambda} = - 2 \kappa \, \mathrm{Tr} \! \left( \delta \Lambda \, \alpha_\phi \right) \mathrm{d}\phi
\end{equation}
and using $\delta \Lambda^{\mathrm{d}} = 0$, no term can combine with the diagonal matrix $D$, so that only the $J_1$ contribution of $\alpha_\phi$ survives. Then, for this reason, only the strictly upper triangular part of the residual symmetries $\Lambda$ can contribute to $q_{\delta \Lambda}$. One therefore finds that, on shell of the flat diagonal gauge \eqref{eq. hs diag gauge},
\begin{equation}
    q_{\delta \Lambda} = - 2 \kappa \, \mathrm{Tr} \! \left( \delta \Lambda^{>} J_1 \right) \mathrm{d}\phi \, .
\end{equation}
Given the peculiar form of this expression, it can be functionally integrated without any assumption on the explicit shape of $\Lambda^{>}$, which a priori may depend on the value of the coupling $s$. This leads us to define generically the Carrollian screening charges as
\begin{equation}\label{screeningcharges}
    S_\Lambda = - 2 \kappa \int \mathrm{d}\phi \, \mathrm{Tr} \! \left( \Lambda^{>} J_1 \right) = - 2 \kappa \, \alpha_s \int \mathrm{d}\phi \, \mathrm{tr} \! \left( \Lambda^{>}_{1} J_1 \right) .
\end{equation}
In particular, we see that the lower triangular part of $\Lambda$ does not appear in any of the symmetry generators, whether asymptotic or pure gauge. The remaining $s(s-1)$ constants that may arise in this sector therefore act as stabilizers of the residual symmetries, as we have already observed for spin two and three. Notice that equation~\eqref{screeningcharges} makes the distinction between our boundary conditions~\eqref{eq. hs diag gauge} and the ones considered in~\cite{Afshar:2016kjj,Ammon:2017vwt} even more explicit. Indeed, in these works, the mode $J_1$ of asymptotically flat spacetimes is not switched on. As a consequence, the screening charges~\eqref{screeningcharges}, which depend on this mode, vanish identically in that case. They therefore do not have access to these quantities, and hence to the corresponding dual interpretation in terms of a Carrollian Coulomb gas.

We now consider a gauge parameter $\Lambda^{(1)}$ which is nontrivial only on the first upper diagonal. It then satisfies
\begin{subequations}
    \begin{align}
        (\Lambda^{(1)}_0)'+[D_0,\Lambda_0^{(1)}] &=0\,,\\
        (\Lambda^{(1)}_1)' + [D_1,\Lambda_0^{(1)}] + [D_0,\Lambda_1^{(1)}] &=0\,.
    \end{align}
\end{subequations}
This is solved by 
\begin{equation}
    \Lambda^{(1)}_0(\phi)+\theta \Lambda_1^{(1)} (\phi) = U(\phi) (c_0^{(1)}+\theta c_1^{(1)}) U(\phi)^{-1}\,,
\end{equation}
where $c_0^{(1)}$ and $c_1^{(1)}$ are constant matrices with all entries zero except for the first upper diagonal, and
\begin{align}
    U(\phi) &= \exp \! \left( -\int^\phi_{\phi_0} \mathrm{d}\phi' \,D(\phi')\right) \\
    &= \exp \! \left( -\int^\phi_{\phi_0} \mathrm{d}\phi' \,D_0(\phi')\right)\left(1-\theta\, \int^\phi_{\phi_0} \mathrm{d}\phi' \,D_1(\phi')\right)\\
    &= U_0(\phi) (1-\theta \,U_1(\phi))\,.
\end{align}
Then, concretely,
\begin{subequations}
    \begin{align}
        \Lambda_0^{(1)} &= U_0\,c_0^{(1)}\,U_0^{-1}\,,\\
        \Lambda_1^{(1)} &= U_0 \,c_1^{(1)}\,U_0^{-1} - U_0\,[U_1,c_0^{(1)}]\,U_0^{-1}\,,
    \end{align}
\end{subequations}
and the screening charge (see~\eqref{screeningcharges}) is
\begin{align}
    S_{\Lambda} &= -2\kappa\,\alpha_s \int d\phi\,\mathrm{tr}(\Lambda_1^{(1)}J_1)\\
    &=-2\kappa\,\alpha_s \int d\phi\, \mathrm{tr}\left( U_0 \,c_1^{(1)}\,U_0^{-1}\,J_1\right) +2\kappa\,\alpha_s \int d\phi\, \mathrm{tr} \left(  U_0\,[U_1,c_0^{(1)}]\,U_0^{-1}\,J_1 \right)\,.
\end{align}
The Poisson brackets of the entries of $D_0$ and $D_1$ with $U_0$ and $U_1$ are
\begin{subequations}
    \begin{align}
        \{ U_0(\phi_1),(D_0)_{ii}(\phi_2)\} & = 0\,,\\
        \{ U_0(\phi_1),(D_1)_{ii}(\phi_2)\} & =\frac{1}{2\kappa \alpha_s} \left(e_{ii}-\frac{1}{s}\right)\,\delta(\phi_1-\phi_2)\,,\\
        \{ U_1(\phi_1),(D_0)_{ii}(\phi_2)\} & = -\frac{1}{2\kappa \alpha_s} \left(e_{ii}-\frac{1}{s}\right)\,\delta(\phi_1-\phi_2)\,,\\
        \{ U_1(\phi_1),(D_1)_{ii}(\phi_2)\} & = 0\,,
    \end{align}
\end{subequations}
where $e_{ii}$ is the diagonal matrix with $1$ in position $i$ and zero otherwise. One can then straightforwardly check that
\begin{subequations}
    \begin{align}
        \{ S_\Lambda, (D_0)_{ii}(\phi)\} &= \mathrm{tr} (e_{ii}[J_1,\Lambda_0^{(1)}])\,,\\
        \{ S_\Lambda, (D_1)_{ii}(\phi)\} &= \mathrm{tr} (e_{ii} [J_1,\Lambda_1^{(1)}])\,,
    \end{align}
\end{subequations}
reproducing~\eqref{gaugetrans D0} and \eqref{gaugetrans D1}. 

The gauge transformations generated by elements supported only on the first upper diagonal give rise, via commutators, to transformations supported on higher upper diagonals. A direct verification that the screening charges generate the corresponding transformations for higher matrices is expected to be subtle, as one must carefully account for a possible field dependence of the integration constants. Nevertheless, no unexpected features are anticipated, since it follows formally from the covariant phase space formalism that the screening charges~\eqref{screeningcharges} coincide with the Noether charges of the pure gauge symmetries~\eqref{eq. Noether pure gauge}, and thus have to satisfy the pure gauge Poisson brackets on the reduced symplectic structure~\eqref{eq. PB hs}.

\subsection{Carrollian Miura transformations}

Consider again the residual gauge transformations parametrized by $\Lambda^>$. We now aim to identify combinations of the entries of $D$ that are invariant under such transformations. An invariant of this type is provided by a differential operator $L$,\footnote{This should not be confused with the Bondi angular momentum aspect of asymptotically flat gravity.} defined as the row-ordered determinant of a matrix-valued operator,
\begin{align}
    L &= \mathrm{rdet} (\partial + D) \\
    &= (\partial + (D_0)_{11}) \cdots (\partial + (D_0)_{ss}) \nonumber\\
    &\quad + \theta \sum_{i=1}^s (\partial + (D_0)_{11}) \cdots (D_1)_{ii} \cdots (\partial + (D_0)_{ss}) \, .
    \label{Miura0}
\end{align}
To see this, we first note that it can be rewritten as
\begin{equation}
    L = \mathrm{rdet}(\partial + J_1 + D) \, .
\end{equation}
An infinitesimal gauge transformation then leads to a modified operator,
\begin{align}
    \partial + J_1 + \tilde{D} &= \partial + J_1 + D + \partial \Lambda + [J_1 + D,\Lambda] + \cdots \\
    &= (1 - \Lambda)(\partial + J_1 + D)(1 + \Lambda) + \cdots \, .
\end{align}
It remains to show that the row-ordered determinant is invariant under such transformations. This is immediate for constant $\Lambda$, but requires a more careful analysis otherwise.

Let $l=\partial+J_1+D$ be a matrix-valued differential operator. Let $P=(P_1,\dots,P_s)^T$ be a column vector of differential operators such that $P_s=1$ and
\begin{equation}\label{differentialsystem_K}
    l P = \begin{pmatrix}
        K\\ 0\\ \vdots \\ 0
    \end{pmatrix} .
\end{equation}
Concretely, this system of equations reads
\begin{subequations}
    \begin{align}
        (\partial + D_{11}) P_1 &= K \, ,\\
        (\partial + D_{ii}) P_i &= P_{i-1} \qquad (i=2,\dots,s)\,.
    \end{align}
\end{subequations}
The last $(s-1)$ equations can be solved recursively, starting from $P_s=1$, and admit a unique solution. The operator $K$ is then uniquely determined by the requirement that~\eqref{differentialsystem_K} admits a solution with $P_s=1$, and one finds
\begin{equation}
    K = (\partial + D_{11}) \cdots (\partial + D_{ss}) = \mathrm{rdet} (\partial + D) = \mathrm{rdet} (\partial + J_1 + D)\,.
\end{equation}

Now consider the transformed operator
\begin{equation}
    \tilde{l}=\partial+ J_1 + \tilde{D} = N (\partial+ J_1 + D) N^{-1}\,,
\end{equation}
for an upper uni-triangular matrix $N$, i.e.\ an upper triangular matrix with unit diagonal (in our case $N=1+\Lambda^>$ with $\Lambda$ infinitesimal). One verifies that $\tilde{P}=NP$ satisfies $\tilde{P}_s = 1$ and
\begin{equation}
    \tilde{l} \tilde{P} = N l P = N \begin{pmatrix}
        K\\0\\ \vdots \\ 0
    \end{pmatrix} = \begin{pmatrix}
        K\\0 \\ \vdots \\ 0
    \end{pmatrix} ,
\end{equation}
where we used that $N$ preserves the first basis vector, i.e.\ $N e_1 = e_1$, which holds for any upper uni-triangular matrix. Hence $\mathrm{rdet}\,\tilde{l}=K$, which implies
\begin{equation}
    (\partial + D_{11}) \cdots (\partial + D_{ss}) = (\partial + \tilde{D}_{11}) \cdots (\partial + \tilde{D}_{ss})\,.
\end{equation}
We now identify the entries of $D$ in terms of fields,
\begin{equation}
    D_{jj} = \frac{1}{\sqrt{\pi \kappa \,\alpha_s}}\,\big(\mathrm{i}\,\vec{\epsilon}_j\cdot \vec{\psi}' + \mathrm{i}\,\theta\,\vec{\epsilon}_j\cdot \vec{\bar{\psi}}'\big)\,,
\end{equation}
where $\vec{\psi}$ and $\vec{\bar{\psi}}$ take values in the weight space of $\mathfrak{sl}(s)$ (identified with $\mathbb{R}^{s-1}$ endowed with the standard scalar product), and the $s$ vectors $\vec{\epsilon}_j$ are the weights of the vector representation of $\mathfrak{sl}(s)$, satisfying
\begin{align}
    \vec{\epsilon}_i\cdot \vec{\epsilon}_j &= \delta_{ij}-\frac{1}{s} \, ,&
    \sum_{j=1}^s \vec{\epsilon}_j &= 0 \, , &
    \sum_{j=1}^s \vec{\epsilon}_j\,{\vec{\epsilon}_j}{}^T &= \mathbf{1} \, .
\end{align}
The invariant differential operator~\eqref{Miura0} can then be expanded in powers of $\partial$,
\begin{subequations}
    \begin{align}
        \big(\alpha \partial + \mathrm{i}\,\vec{\epsilon}_1\cdot \vec{\psi}'\big) \cdots \big(\alpha \partial + \mathrm{i}\,\vec{\epsilon}_s\cdot \vec{\psi}'\big)
        &= (\alpha \partial)^s + \sum_{j=2}^s u_j \,(\alpha \partial)^{s-j} \, ,\\
        \sum_{j=1}^s \big(\alpha \partial + \mathrm{i}\,\vec{\epsilon}_1\cdot \vec{\psi}'\big) \cdots \big(\mathrm{i}\,\vec{\epsilon}_j\cdot \vec{\bar{\psi}}'\big)\cdots \big(\alpha\partial + \mathrm{i}\,\vec{\epsilon}_s\cdot \vec{\psi}'\big)
        &= \sum_{j=2}^s \bar{u}_j \,(\alpha \partial)^{s-j} \, ,
    \end{align}
\end{subequations}
with $\alpha = \sqrt{\pi \kappa \,\alpha_s}$.
The coefficient fields $u_j$ and $\bar{u}_j$ are combinations of the fields $\vec{\psi}$ and~$\vec{\bar{\psi}}$ that are invariant under residual gauge transformations. The above relation is thus the Carrollian analogue of the Miura transformation, and it coincides (up to normalization factors) with the expression in~\cite{Fredenhagen:2025aqd} in the case where one of the central charges vanishes,\footnote{In~\cite{Fredenhagen:2025aqd}, this corresponds to $\alpha^-=0$.}~$c_1=0$.


\section{Conclusion} \label{sec. conclu}

In this work we have provided a bulk holographic realization, in three-dimensional asymptotically flat spacetimes coupled to higher-spin fields, of the Carrollian free-field structure of the generalized higher-spin $\mathfrak{bms}_3$ algebra at null infinity. To this end, we adopted the ``flat from AdS'' perspective~\cite{Campoleoni:2023fug}, in which the holographic gauge choice and solution space of the flat case are obtained from their counterparts of asymptotically AdS spacetimes via a Laurent expansion in the cosmological constant. Within this framework, we began by reviewing the construction of the diagonal gauge in AdS~\eqref{eq. AdS diag gauge} in the Chern--Simons formulation~\cite{BALOG199076,Campoleoni:2017xyl}, both in the purely spin-two case and in the presence of higher-spin couplings.

In the standard gauge used in (higher-spin) AdS/CFT, namely the Drinfeld--Sokolov gauge~\eqref{eq. AdS DS gauge}, the selected symplectic slice at conformal infinity coincides with the fully reduced phase space, in which all degenerate directions have been eliminated. In this setting and for the gauge algebra $\mathfrak{sl}(s)\times \mathfrak{sl}(s)$, the asymptotic symmetries are described by two copies of the $\mathcal{W}_s$-algebra. In contrast, the diagonal gauge restores pure gauge symmetries within the residual symmetry algebra, yielding a simpler, Abelian symplectic structure, albeit with a nontrivial residual kernel. The asymptotic symmetries are then expressed in terms of free fields, while the physical $\mathcal{W}$-observables arise as equivalence classes modulo the residual kernel. These observables are captured by the Miura transformations~\eqref{eq. AdS Miura}, which provide the gauge-invariant combinations of the currents. The boundary theory is thus governed by the classical Coulomb gas formalism~\cite{Fateev:1987zh}. In particular, the generators of the restored gauge symmetries were shown in~\cite{Campoleoni:2017xyl} to coincide with the corresponding screening charges. We further extended this result within the covariant phase space formalism~\cite{Lee:1990nz,Wald:1993nt,Iyer:1994ys} by demonstrating that these generators admit a more fundamental interpretation as genuine Noether charges~\eqref{eq. qLambda} associated with pure gauge transformations, thereby distinguishing them from the asymptotic Iyer--Wald charges~\eqref{eq. kqqdelta}. This additional insight proved particularly useful for the asymptotically flat analysis, where the structure of the dual theory remains much less established.

We emphasize that we worked in radial gauge~\eqref{eq. radialgauge}, in which both the on-shell action and the surface charges are automatically finite at asymptotic infinity, without requiring holographic renormalization. This suggests that such symmetry algebras can, in principle, be realized from the bulk on any finite radial slice~\cite{Compere:2014cna}. It would be worthwhile to further investigate this aspect of the radial gauge along the lines of~\cite{Grumiller:2016pqb,Ciambelli:2023ott}, as well as its implications for boundary and corner symplectic prescriptions~\cite{Freidel:2019ohg,McNees:2023tus,Campoleoni:2023eqp,Riello:2024uvs,Campoleoni:2025bhn}. Moreover, we observed that, upon radial reconstruction, the metric associated with the diagonal Chern--Simons gauge takes a form corresponding to the Weyl--Fefferman--Graham relaxation~\eqref{eq. WFG gauge}, whereas the Drinfeld--Sokolov gauge is metrically realized in terms of the standard Fefferman--Graham gauge~\eqref{eq. FG gauge}. It turns out that this standard rewriting comes at the price of a more intricate structure, as the reduced symplectic form is given by the inverse of the Virasoro bracket. These observations further support the idea that a unified treatment of these two perspectives within a generalized Weyl--Fefferman--Graham framework, explored for spin two in~\cite{Ciambelli:2023ott} and for higher spins in~\cite{Delfante:2025lxn}, constitutes a promising direction toward a deeper quantum understanding of holographic duality. We leave this investigation for future work, noting that it becomes even more compelling in the case of vanishing cosmological constant, where such structures remain largely unexplored.

We then generalized the diagonal gauge to asymptotically flat spacetimes by implementing Carrollian prescriptions for the fields~\eqref{eq. flat s=3 flat prescriptions}, leading to the gauge configuration~\eqref{eq. hs boson gauge}. Interestingly, this construction matches the more algebraic approach of~\eqref{eq. hs diag gauge}, thereby confirming that this gauge admits a clear geometric interpretation. In particular, it corresponds to a gauge in which the degrees of freedom are not switched on along the lowest-weight modes, as in the DS~\eqref{eq. AdS DS gauge} or BMS~\eqref{eq. flat Bondi gauge} gauges, but instead along diagonal generators. As a result, much like its asymptotically AdS counterpart, this asymptotically flat diagonal gauge exhibits a simpler reduced symplectic structure, now expressed in terms of Carrollian scalar fields. An interesting open question is whether this symplectic structure can be obtained from an electric or magnetic limit of its Lorentzian ancestor, an avenue that is relevant both classically and quantum mechanically.

This construction leads to a realization of flat asymptotic symmetries in terms of Carrollian free fields, achieved through the restoration of pure gauge symmetries in comparison with the standard Bondi gauge. The corresponding generators are again identified, within the CPS formalism, as Noether charges~\eqref{eq. flat screening}, and can be interpreted at the boundary as Carrollian counterparts of screening charges. This suggests that the dual theory may be governed by a Carrollian analogue of the Coulomb gas formalism familiar from the AdS case, thereby providing a possible route toward the quantization of flat holography. The physical observables, namely the generalized higher-spin mass and angular momentum aspects, are identified through the Carrollian Miura transformations~\eqref{eq. flat Carroll Miura}. These were originally introduced from a purely field-theoretic perspective~\cite{Banerjee:2015kcx,Fredenhagen:2025aqd} as the combinations realizing the higher-spin $\mathfrak{bms}_3$ algebra in terms of Carrollian free fields~\eqref{eq. flat bosonic charge algebra}. Here, we have provided a holographic realization of these structures in the context of higher-spin gravity in asymptotically flat spacetimes, through the flat diagonal gauge in the CS formulation. This result is particularly significant, as the resulting reduced symplectic structure, corresponding to that of a Carrollian scalar field, is well suited for quantizing the dual Carrollian theory. Its holographic emergence thus constitutes a promising step toward a concrete realization of a (higher-spin) BMS/CFT correspondence.

At the metric level, the change of symplectic slice from~\eqref{eq. flat Bondi gauge} to~\eqref{eq. flat bosonic gauge} induces a transition from the standard Bondi gauge~\eqref{eq. Bondi gauge} to a Bondi--Weyl type gauge~\eqref{eq. Bondi Weyl gauge}. This observation is suggestive for future developments. Actually, another class of relaxations of the Bondi gauge, known as the covariant Bondi gauge, has been studied in~\cite{Campoleoni:2022wmf} for spin two and in~\cite{Delfante:2025lxn} for higher spins. These works showed that such relaxations allow for a clear identification of sources and vacuum expectation values of the dual Carrollian theory directly from the bulk expansion. Understanding the interplay between these different relaxations appears to be a promising avenue toward a well-defined formulation of (higher-spin) holography in flat space. A further direction would be to explore a generalization of the fluid/gravity correspondence~\cite{Campoleoni:2018ltl,Campoleoni:2022wmf} to higher-spin theories. In this framework, the bulk flat limit maps the boundary theory from a relativistic to a Carrollian hydrodynamic system. The understanding of how higher-spin couplings deform this transition may provide new insights into both the structure of the dual field theory and fluid dynamics.

We conclude by outlining several additional directions for future work. Beyond the Weyl--Fefferman--Graham and Bondi--Weyl relaxations, it would be interesting to relate the Carrollian diagonal gauge to other gauge choices and boundary conditions for asymptotically flat spacetimes explored in the literature (see, e.g.,~\cite{Detournay:2016sfv,Afshar:2016kjj,Grumiller:2017sjh,Adami:2020ugu,Adami:2022ktn,Adami:2024rkr,Taghiloo:2024ewx}), where Carrollian Heisenberg algebras have also been discussed. It could furthermore be worthwhile to investigate possible connections with integrable systems, in which configurations with fields turned on along diagonal algebraic directions naturally arise~\cite{Fuentealba:2017omf,Ojeda:2020bgz}. Incorporating chemical potentials in the flat diagonal gauge would also make it possible to relate to (higher-spin) flat space cosmologies~\cite{Bagchi:2012xr,Barnich:2012xq,Gary:2014ppa}, in close analogy with the AdS case, where such deformations allow for the description of (higher-spin) black holes in three dimensions~\cite{Gutperle:2011kf,Bunster:2014mua}. It could therefore be illuminating to clarify how the soft hair structure maps onto the Carrollian Coulomb gas formalism and its associated screening charges.


\section*{Acknowledgments}

We dedicate this work to the memory of Robert G. Leigh. We would like to thank Andrea Campoleoni, Adrien Fiorucci, Daniel Grumiller, Chrysoula Markou, Lea Mele and Stefan Prohazka for useful discussions.


\bibliographystyle{JHEP}

\providecommand{\href}[2]{#2}
\end{document}